\documentclass{nature}

\usepackage{xspace}
\usepackage{color}
\usepackage{graphics}
\usepackage{epsfig}
\usepackage{upgreek}
\usepackage{amssymb}
\usepackage{amsmath}
\usepackage{siunitx}
\usepackage{booktabs}
\usepackage{mathrsfs}
\usepackage{url}
\usepackage{caption}

\usepackage{lineno}

\usepackage{xcolor}
\definecolor{orange}{HTML}{FFA500}
\definecolor{cyan}{HTML}{48D1CC}
\definecolor{purple}{HTML}{8D32C1}
\definecolor{green}{HTML}{008000}

\def\frb{FRB 180916.J0158+65\xspace}

\def\Halpha{H$\alpha$\xspace}
\def\dm{348.76~pc~cm$^{-3}$\xspace}

\def\dmu{pc~cm$^{-3}$\xspace}

\newcommand\ion[2]{\text{#1\,\textsc{\lowercase{#2}}}}

\def\pasa{Pubs.\ Astron.\ Soc.\ Australia}
\def\nat{Nature}
\def\aap{Astron.\ \& Astrophys.}
\def\aaps{Astron.\ \& Astrophys.\ Suppl.\ Ser.}
\def\apj{Astrophys.\ J.}
\def\apjs{Astrophys.\ J.\ Suppl.}
\def\apjl{Astrophys.\ J. Letters}
\def\aj{Astron.\ J.}
\def\mnras{Mon.\ Not.\ R.\ Astron.\ Soc.}
\def\aapr{Astron.\ \& Astrophys.\ Rev.}
\def\pasp{Publ.\ Astron.\ Soc.\ Pac.}
\def\raa{Res.\ in Astron.\ \& Astrophys.}
\newcommand{\araa}{Ann. Rev. Astron. \& Astrophys.}

\graphicspath{ {./} }

\title{A repeating fast radio burst source localised to a nearby spiral galaxy}

\author{B.~Marcote$^{1}$, K.~Nimmo$^{2,3}$, J.~W.~T.~Hessels$^{2,3}$, S.~P.~Tendulkar$^{4,5}$, C.~G.~Bassa$^{2}$, Z.~Paragi$^{1}$, A.~Keimpema$^{1}$, M.~Bhardwaj$^{4,5}$, R.~Karuppusamy$^{6}$, V.~M.~Kaspi$^{4,5}$, C.~J.~Law$^{7}$, D.~Michilli$^{4,5}$, K.~Aggarwal$^{8,9}$, B.~Andersen$^{4,5}$, A.~M.~Archibald$^{10,3}$, K.~Bandura$^{11,9}$, G.~C.~Bower$^{12}$, P.~J.~Boyle$^{4,5}$, C.~Brar$^{4,5}$, S.~Burke-Spolaor$^{8,9}$, B.~J.~Butler$^{13}$, T.~Cassanelli$^{14,15}$, P.~Chawla$^{4,5}$, P.~Demorest$^{13}$, M.~Dobbs$^{4,5}$, E.~Fonseca$^{4,5}$, U.~Giri$^{16,17}$, D.~C.~Good$^{18}$, K.~Gourdji$^{3}$, A.~Josephy$^{4,5}$, A.~Yu.~Kirichenko$^{19,20}$, F.~Kirsten$^{21}$, T.~L.~Landecker$^{22}$, D.~Lang$^{16}$, T.~J.~W.~Lazio$^{23}$, D.~Z.~Li$^{24,25}$, H.-H.~Lin$^{25}$, J.~D.~Linford$^{13}$, K.~Masui$^{26,27}$, J.~Mena-Parra$^{26}$, A.~Naidu$^{4,5}$, C.~Ng$^{14}$, C.~Patel$^{4,5}$, U.-L.~Pen$^{25,28,14,16}$, Z.~Pleunis$^{4,5}$, M.~Rafiei-Ravandi$^{16}$, M.~Rahman$^{14}$, A.~Renard$^{14}$, P.~Scholz$^{14,22}$, S.~R.~Siegel$^{4,5}$, K.~M.~Smith$^{16}$, I.~H.~Stairs$^{18}$, K.~Vanderlinde$^{14,15}$ \& A.~V.~Zwaniga$^{4,5}$}

\begin{document}

\maketitle

\begin{affiliations}
    \item Joint Institute for VLBI ERIC (JIVE), Oude Hoogeveensedijk 4, 7991~PD Dwingeloo, The Netherlands.
    \item ASTRON, Netherlands Institute for Radio Astronomy, Oude Hoogeveensedijk 4, 7991~PD Dwingeloo, The Netherlands.
    \item Anton Pannekoek Institute for Astronomy, University of Amsterdam, Science Park 904, 1098~XH Amsterdam, The Netherlands.
    \item Department of Physics, McGill University, 3600 University Street, Montr\'{e}al, Qu\'{e}bec H3A~2T8, Canada.
    \item McGill Space Institute, McGill University, 3550 University Street, Montr\'{e}al, Qu\'{e}bec H3A~2A7, Canada.
    \item Max-Planck-Institut f\"ur Radioastronomie, Auf dem H\"{u}gel 69, D-53121 Bonn, Germany.
    \item Department of Astronomy and Owens Valley Radio Observatory, California Institute of Technology, Pasadena, California 91125, USA.
    \item Department of Physics and Astronomy, West Virginia University, P.O. Box 6315, Morgantown, West Virginia 26506, USA.
    \item Center for Gravitational Waves and Cosmology, West Virginia University, Chestnut Ridge Research Building, Morgantown, West Virginia 26505, USA.
    \item School of Mathematics, Physics, and Statistics, University of Newcastle, Newcatle upon Tyne, NE1~7RU, United Kingdom.
    \item Lane Department of Computer Science and Electrical Engineering, Morgantown, West Virginia 26506, USA.
    \item Academia Sinica Institute of Astronomy and Astrophysics, 645 N. A'ohoku Place, Hilo, Hawaii 96720, USA.
    \item National Radio Astronomy Observatory, Socorro, New Mexico 87801, USA.
    \item Dunlap Institute for Astronomy \& Astrophysics, University of Toronto, 50 St. George Street, Toronto, Ontario M5S~3H4, Canada.
    \item Department of Astronomy \& Astrophysics, University of Toronto, 50 St. George Street, Toronto, Ontario M5S~3H4, Canada.
    \item Perimeter Institute for Theoretical Physics, Waterloo, Ontario N2L~2Y5, Canada.
    \item Department of Physics and Astronomy, University of Waterloo, Waterloo, Ontario N2L~3G1, Canada.
    \item Department of Physics and Astronomy, University of British Columbia, 6224 Agricultural Road, Vancouver, British Columbia V6T~1Z1, Canada.
    \item Instituto de Astronom\'{i}a, Universidad Nacional Aut\'{o}noma de M\'{e}xico, Apdo. Postal 877, Ensenada, Baja California 22800, M\'{e}xico.
    \item Ioffe Institute, 26 Politekhnicheskaya st., St. Petersburg 194021, Russia.
    \item Department of Space, Earth and Environment, Chalmers University of Technology, Onsala Space Observatory, 439~92, Onsala, Sweden.
    \item Dominion Radio Astrophysical Observatory, Herzberg Astronomy \& Astrophysics Research Centre, National Reseach Council Canada, P.O. Box 248, Penticton, British Columbia V2A~6J9, Canada.
    \item Jet Propulsion Laboratory, California Institute of Technology, Pasadena, California 91109, USA.
    \item Department of Physics, University of Toronto, 60 St. George Street, Toronto, Ontario M5S~1A7, Canada.
    \item Canadian Institute for Theoretical Astrophysics, 60 St. George Street, Toronto, Ontario M5S~3H8, Canada.
    \item MIT Kavli Institute for Astrophysics and Space Research, Massachusetts Institute of Technology, 77 Massachusetts Ave., Cambridge, Massachusetts 02139, USA.
    \item Department of Physics, Massachusetts Institute of Technology, Cambridge, 77 Massachusetts Ave., Massachusetts 02139, USA.
    \item Canadian Institute for Advanced Research, CIFAR Program in Gravitation and Cosmology, Toronto, Ontario M5G 1Z8, Canada.
\end{affiliations}

\begin{abstract}

Fast radio bursts (FRBs) are brief, bright, extragalactic radio flashes\cite{lorimer2007,petroff2019}.
Their physical origin remains unknown, but dozens of possible models have been postulated\cite{platts2018}.
Some FRB sources exhibit repeat bursts\cite{spitler2016,chime2019b,chime2019c,kumar2019}.
Though over a hundred FRB sources have been discovered to date\cite{petroff2016}, only four have been localised and associated with a host galaxy\cite{chatterjee2017,ravi2019,bannister2019,prochaska2019}, with just one of the four known to repeat\cite{chatterjee2017}.
The properties of the host galaxies, and the local environments of FRBs, provide important clues about their physical origins.  
However, the first known repeating FRB has been localised to a low-metallicity, irregular dwarf galaxy, and the apparently non-repeating sources to higher-metallicity, massive elliptical or star-forming galaxies, suggesting that perhaps the repeating and apparently non-repeating sources could have distinct physical origins.
Here we report the precise localisation of a second repeating FRB source\cite{chime2019c}, \frb, to a star-forming region in a nearby (redshift \boldmath{$z = 0.0337 \pm 0.0002$}) massive spiral galaxy, whose properties and proximity distinguish it from all known hosts.
The lack of both a comparably luminous persistent radio counterpart and a high Faraday rotation measure\cite{chime2019c} further distinguish the local environment of \frb from that of the one previously localised repeating FRB source, FRB~121102.  
This demonstrates that repeating FRBs have a wide range of luminosities, and originate from diverse host galaxies and local environments.

\end{abstract}

The CHIME/FRB Collaboration is beginning to discover many repeating FRB sources\cite{chime2019b,chime2019c}, which allows subsequent targeted observations using distributed radio telescope arrays to obtain precise interferometric localisations.  CHIME/FRB discovered\cite{chime2019c} the repeating source \frb, which we observed at a central frequency of 1.7~GHz and bandwidth of 128~MHz using eight radio telescopes of the European Very-long-baseline-interferometry Network (EVN) for 5.5 hours on June 19th, 2019. As described in the Methods, we simultaneously recorded both EVN single-dish raw voltage data as well as high-time-resolution intensity data using the PSRIX data recorder\cite{lazarus2016} in filterbank mode at the 100-m Effelsberg telescope.

\begin{figure}
   \centerline{\resizebox{\textwidth}{!}{\includegraphics{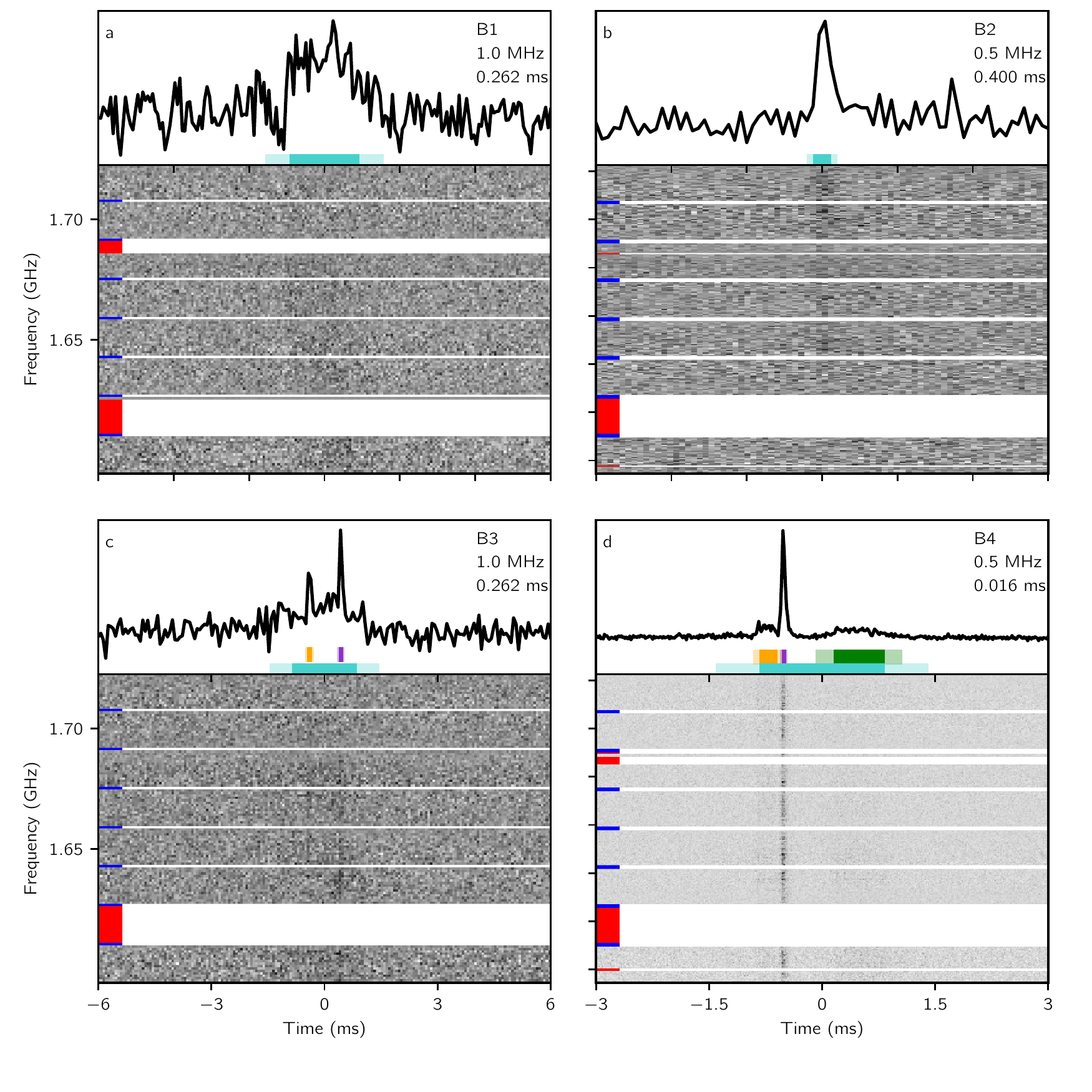}}}
    \caption{\label{fig:efauto_profile}
    {\bf Burst detections in Effelsberg auto-correlation data.}
    Band-averaged profiles and dynamic spectra of the four bursts, as detected using the coherently dedispersed Effelsberg auto-correlation data ({\bf a}, {\bf b}, {\bf c} and {\bf d}). Each burst was fitted with a Gaussian distribution to determine the full-width at half-maximum (FWHM) durations, which are represented by the dark cyan bars. The lighter cyan encloses the 2-$\sigma$ region.
    Bursts B3 and B4 ({\bf c} and {\bf d}) show sub-bursts; the orange, purple and green bars correspond to the FWHM of these sub-bursts, and the lighter bar (of each colour) encloses the 2-$\sigma$ region. Note the different time windows plotted: B1 and B3 ({\bf a} and {\bf c}) show 12~ms, whereas B2 and B4 ({\bf b} and {\bf d}) show 6~ms surrounding the burst peak. The solid white lines represent frequency channels that have been removed from the data due to either RFI or subband edges, indicated by the red and blue markers, respectively. The burst data have been downsampled by various factors in time and frequency to optimise the visual representation. The RFI excision was done prior to downsampling. The time and frequency resolution used for plotting is shown in the top right of each panel.}
\end{figure}

In a search of the PSRIX data, we detected four bursts from \frb with signal-to-noise ratios between 9.5 and 46. The observed dispersion measures (DM) of the bursts are consistent with those previously reported\cite{chime2019c} for this source.  The burst properties, as derived from the PSRIX data, are listed in Extended Data Table~\ref{tab:detection_properties}. No other dispersed single pulses of plausible astrophysical origin were found in this search, for DMs in the range $0$--$700$~pc~cm$^{-3}$. Using the EVN raw voltage data, we generated high-time-resolution (16-$\upmu$s samples) Effelsberg auto-correlation data containing each burst. We used coherent dedispersion to mitigate the intra-channel smearing that dominated over the temporal resolution in the PSRIX data. By minimising dispersion broadening, we properly resolve the burst structure, and find a best-fit DM $= 348.76 \pm 0.10$~pc~cm$^{-3}$ using the brightest burst (Methods).  We also detect brightness modulation with a characteristic frequency scale of $59 \pm 13$~kHz, which we interpret as scintillation imparted by the ionised interstellar medium of the Milky Way (Methods).  The properties of the four bursts, as seen in the Effelsberg auto-correlations, are also shown in Extended Data Table~\ref{tab:detection_properties}. The frequency-averaged burst profiles and dedispersed dynamic spectra are shown in Figure~\ref{fig:efauto_profile} and Extended Data Figure~\ref{fig:efpsr_profile} for the Effelsberg auto-correlation and PSRIX data, respectively.

\begin{figure}
   \centerline{\resizebox{!}{0.75\textheight}{\includegraphics{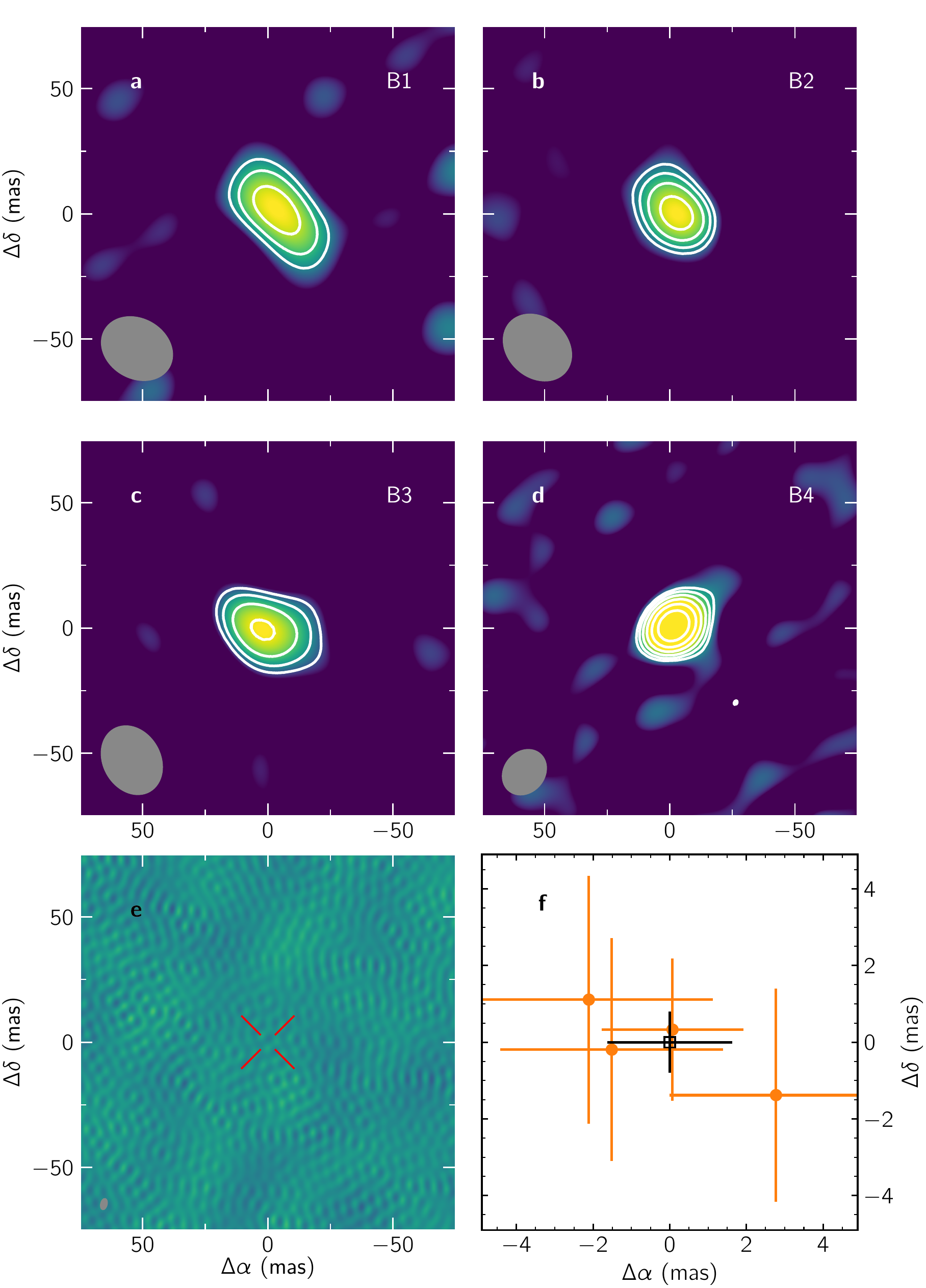}}}
    \caption{\label{fig:evn-bursts}
    {\bf EVN images and burst positions.} 
    Each individual burst is imaged by using the EVN gated visibility data ({\bf a}, {\bf b}, {\bf c} and {\bf d}).  Contours start at three times the r.m.s. noise level of each image (75, 65, 43, 100~mJy~beam$^{-1}$, and 9.7~$\upmu$Jy~beam$^{-1}$, respectively) and increase by factors of $\sqrt{2}$. {\bf e:} image using the full span of the EVN observation and 2-s integration time; no significant ($>3\sigma$) persistent radio emission is detected at the positions of the bursts (denoted by the red cross) and no signal above the 4-$\sigma$ level is detected anywhere within the full field.  For all images, the synthesised beam is represented by the grey ellipse in the bottom-left corner.  The Tianma station was only included in the derivation of the continuum image ({\bf e}).  {\bf f:} the positions derived from each individual burst and their associated 1-$\sigma$ uncertainties (orange), with respect to the weighted-average burst position (black). All positions are referred to $\alpha = 01^\text{h}58^\text{m}00.7502^\text{s},\ \delta = 65^\circ 43^\prime 00.3152^{\prime\prime}$.}
\end{figure}

We then used the EVN raw voltage data to create coherently dedispersed cross-correlations, also known as visibilities, at the times of the four bursts.  Radio interferometric images with milli-arcsecond resolution were produced for each individual burst, using a single 0.7--2.7-ms integration, depending on the burst duration. Each image shows emission above at least seven times the r.m.s. noise level (Figure~\ref{fig:evn-bursts}; see Methods for a detailed explanation of the calibration and imaging process).
The four burst images provide an average J2000 position for \frb of $\alpha = 01^\text{h}58^\text{m}00.7502^\text{s} \pm 2.3\ \text{mas},\ \delta = 65^\circ 43^\prime 00.3152^{\prime\prime} \pm 2.3\ \text{mas}$.
Visibilities with 2-second time resolution were also generated for the entire observation span.  We used these to produce an image of the field around \frb in order to search for a persistent radio counterpart, like that seen in the case of FRB~121102\cite{chatterjee2017,marcote2017}. No such emission is detected above a 3-$\sigma$ r.m.s. noise level of 30~$\upmu$Jy~beam$^{-1}$.  Independent observations using the Karl G.\ Jansky Very Large Array (VLA) at 1.6~GHz detect no coincident emission above a 3-$\sigma$ r.m.s. noise level of 18~$\upmu$Jy~beam$^{-1}$ (see Methods and Extended Data Figure~\ref{fig:vla-map}).

The precise EVN position shows that \frb is spatially coincident with a galaxy catalogued in the Sloan Digital Sky Survey\cite{alam2015} as SDSS~J015800.28+654253.0.  Given the maximum expected redshift of approximately 0.11, as inferred by the measured DM (Methods), we find that the probability for chance coincidence is less than 1\% for any type of galaxy with mass greater than $\sim$ $40$\% that of the FRB~121102 host (Methods). Moreover,  we thus conclude that the association of \frb to this particular galaxy is significant.
\frb is close to the plane of the Milky Way, with Galactic longitude $l = 129.7^{\circ}$ and latitude $b = 3.7^{\circ}$.  Its low DM excess compared to the Milky Way contribution (Methods) brought into question whether it could possibly be a Galactic disk or halo object\cite{chime2019c}, but the host galaxy association shows that it is clearly extragalactic.

\begin{figure}
	\centerline{\resizebox{0.8\textwidth}{!}{\includegraphics{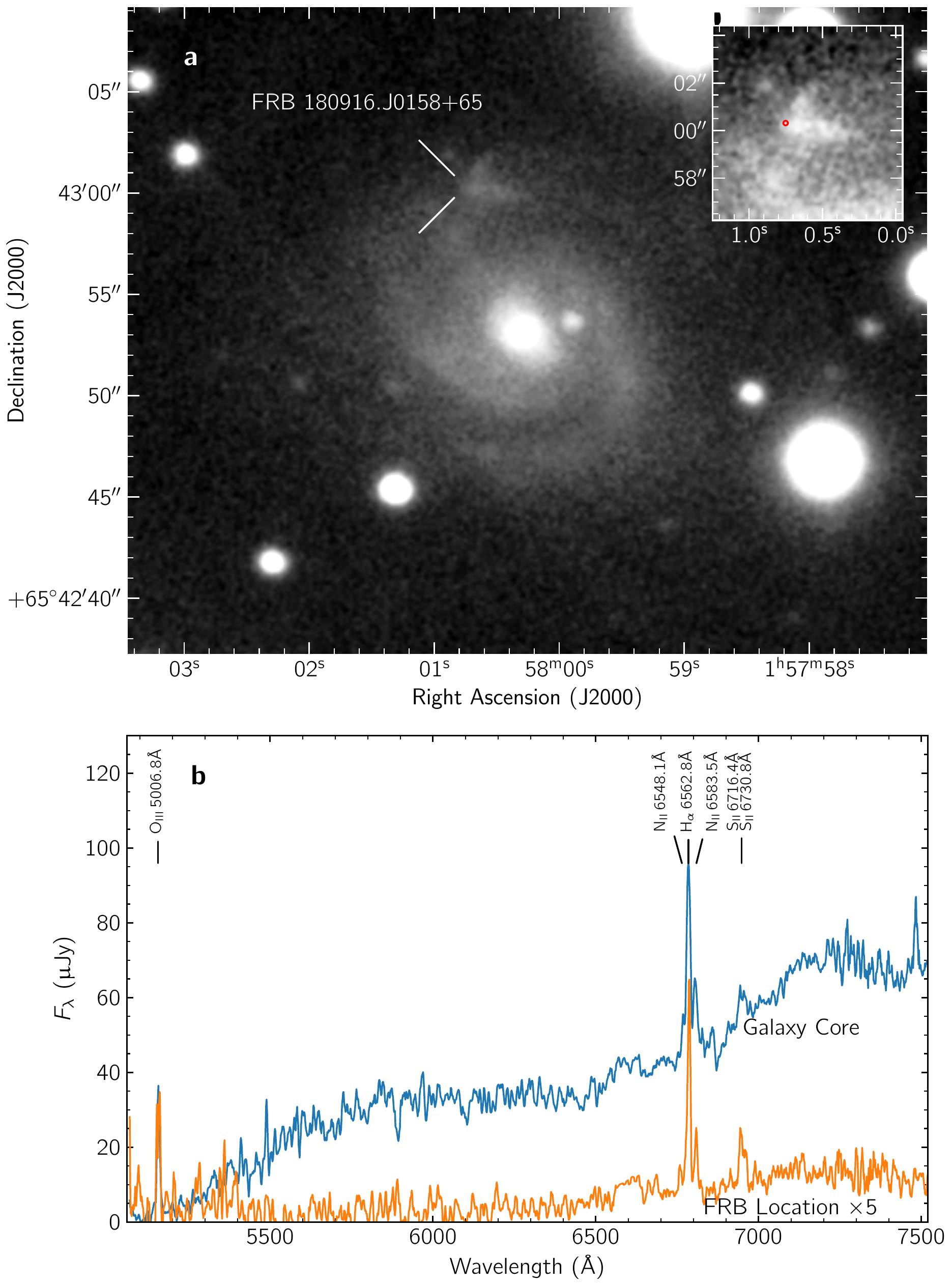}}}
	\caption{\label{fig:galaxy} \label{fig:galaxy_spectra}
	{\bf Gemini-North host galaxy image and optical spectrum.}
	{\bf a:} image of the host galaxy using the $r^{\prime}$ filter.  The position of \frb is marked. The inset shows a higher contrast zoom-in of the star-forming region containing \frb (marked by the red circle). The uncertainty in the position of \frb is smaller than the resolution of the image.  {\bf b:} sky-subtracted spectrum extracted from a 5~arcsec aperture around the host galaxy core (blue) and a 2~arcsec aperture around the location of \frb (orange, scaled by a factor of five for clarity). Emission lines are identified along with their rest-frame wavelengths in air. Due to the complicated shape of the galaxy, the fluxes have not been corrected for slit losses.}
\end{figure}

We used the 8-m Gemini-North telescope to characterise the morphology of the host galaxy and to measure a spectroscopic redshift (Methods).  Deep optical imaging reveals that the host is a nearly face-on spiral galaxy (Figure~\ref{fig:galaxy}) with a total stellar mass of approximately $10^{10}$ times that of the Sun (Methods), which is comparable to the total stellar mass of the Milky Way.  No other comparably large and bright galaxy is visible in the broader field covered by Gemini-North (Extended Data Figure~\ref{fig:galaxy-wide-FoV}).  The milli-arcsecond precision of the EVN localisation shows that the FRB source is close to a bright feature in $r^\prime$-band approximately 7~arcsec (projected separation of roughly 4.7~kpc) from the core of the host galaxy.

With Gemini-North long-slit spectroscopy, we simultaneously targeted the host galaxy centre and the offset position of \frb (Extended Data Figure~\ref{fig:galaxy}).  This revealed strong \Halpha emission, and several other spectral lines commonly associated with star formation (Figure~\ref{fig:galaxy_spectra}).  By measuring optical line ratios, we confirm that the host is indeed a star-forming galaxy (Extended Data Figure~\ref{fig:bpt}).

Comparing with the rest frequencies of the lines, we find a redshift $z = 0.0337 \pm 0.0002$, which corresponds to a luminosity distance of $149.0 \pm 0.9$~Mpc (or an angular size distance of $139.4 \pm 0.8$~Mpc) using standard cosmological parameters\cite{wright2006}.  \frb is thus the closest-known FRB source with a robust host galaxy identification and measured redshift.  It is a factor of six closer than the repeater FRB~121102\cite{tendulkar2017}, and more than an order-of-magnitude closer than the (thus far) non-repeating sources FRB~180924\cite{bannister2019}, FRB~181112\cite{prochaska2019} and FRB~190523\cite{ravi2019}.

The optical spectrum at the location of \frb shows that both the associated star-forming clump and the host galaxy core are at the same redshift (Figure~\ref{fig:galaxy_spectra}).  However, the spectrum at the location of \frb\ is dominated by emission from the clump, which is offset from the centre of the slit. Hence we have no independent direct estimate of the emission or dispersion measure at the position of \frb from the Gemini-North data.  By considering the various modelled foreground contributions to DM, we estimate that the host contribution is less than approximately 70~\dmu, and could be substantially smaller (Methods).

\frb is located at the apex of the `v-shaped' star-forming clump (Extended Data Figure~\ref{fig:galaxy-zoom}) with a relatively large\cite{gusev2014} projected size of roughly $1.5$~kpc and a star-formation rate of $\gtrsim 0.016\ \text{M}_\odot\ \text{yr}^{-1}$ (and a star-formation surface density of $\approx 10^{-2}\ \text{M}_\odot\ \text{yr}^{-1}\ \text{kpc}^{-2}$). The v-shaped star-forming clump is a remarkable feature of the galaxy, suggesting the possibility that the region has undergone an interaction that triggered the star formation, either between multiple star-forming regions or conceivably involving a putative dwarf satellite companion.

The spiral host galaxy of \frb contains over 100 times more stellar mass and has five times higher metallicity (see Extended Data Figure~\ref{fig:bpt} and Methods) than the dwarf host galaxy of FRB~121102.  The discovery of this host demonstrates that some FRB sources exist in galaxies more similar to our own Milky Way.  Previously, it has been noted\cite{tendulkar2017,marcote2017,metzger2017} that the host of FRB~121102 is similar to the type of low-metallicity galaxies with high specific star-formation rate that are associated with hydrogen-poor superluminous supernovae and long-duration gamma-ray bursts.  In contrast, the \frb host is unlike such galaxies; this weakens the case for a general link between all repeating FRB sources and these extreme astrophysical explosions.  In a search of the Open Supernova Catalog\cite{guillochon2017}, we find no previous supernovae or gamma-ray bursts at the location of \frb.

The proximity of \frb constrains the presence of any persistent radio counterpart to a luminosity (at a 3-$\sigma$ confidence level) of $\nu L_\nu < 1.3 \times 10^{36}$ erg~s$^{-1}$ from the continuum EVN data (sensitive to milliarcsecond scales) and to $\nu L_\nu < 7.6 \times 10^{35}$ erg~s$^{-1}$ from the VLA data (sensitive to emission on arcsecond scales; see Methods). Compared to the persistent source associated with the repeating FRB~121102, this upper limit implies that any such source associated with \frb must be at least 400 times fainter than the one associated with FRB~121102\cite{marcote2017}. The previously determined\cite{chime2019c} Faraday rotation measure of \frb, RM~$= -114.6 \pm 0.6$~rad~m$^{-2}$, is three orders of magnitude lower than that of FRB~121102\cite{michilli2018}, where RM~$\sim 10^5$~rad~m$^{-2}$.  As previously suggested\cite{chime2019c}, we conclude that \frb is located in a much less extreme local environment compared with FRB~121102, and that the physical mechanism for FRB repetition does not depend on such conditions.  Nonetheless, models originally proposed for FRB~121102\cite{margalit2018,metzger2019} --- in which the bursts originate from a young and rapidly rotating magnetar --- could potentially still explain the observed properties of \frb by invoking an age of $\sim$300~yr, which is ten times older than that proposed for FRB~121102 (see Methods for a brief comparison to existing models).

While the host galaxies of \frb and FRB~121102 are markedly different, both sources are located near or within a star-forming region in the host galaxy.  This contrasts with the elliptical host galaxies of FRB~180924\cite{bannister2019} and FRB~190523\cite{ravi2019}, where there is comparatively little active star formation, but may be consistent with the star-forming galaxy of FRB~181112\cite{prochaska2019}. This diversity in hosts and local environments allows for the possibility that repeating and apparently non-repeating FRB sources have physically distinct origins.  However, comparison of FRB event rates with those of proposed progenitors disfavours models that invoke cataclysmic explosions and suggests that a large fraction of sources must be capable of repeating\cite{ravi2019b}.  The recent finding that FRB~171019 produces repeat bursts that are almost 600 times fainter compared to the originally discovered signal\cite{kumar2019} underscores the fact that the detectability of repetition depends on instrumental sensitivity and source proximity.  If \frb were at the distance of the other well-localised FRBs, only a small fraction of its (brightest) bursts would be visible.

Furthermore, it has been proposed that a young magnetar origin for the bursts of as-yet non-repeating FRBs in non-star-forming regions is still viable as long as it is possible to form such sources through a variety of channels, including direct stellar collapse, accretion-induced collapse, and through the merger of compact objects\cite{margalit2019}.  Ultimately, a larger number of precision localisations is needed before we can establish that multiple physical origins are required to explain the observed FRB phenomenon.  There are now 11 repeating FRBs known\cite{spitler2016,chime2019b,kumar2019,chime2019c}, and with precision localisations it will be possible to establish whether repeating and apparently non-repeating FRB sources have demonstrably different environments.

FRBs have now been localised with luminosity distances that span approximately 150~Mpc to 4~Gpc.  Estimating distance purely based on DM, it appears likely that there are FRBs that are even closer\cite{mahony2018} or more distant\cite{bhandari2018,chime2019c} compared with this range.  The four bursts presented here have isotropic-equivalent spectral energy densities as low as approximately $5\times10^{27}$~erg~Hz$^{-1}$ (Table~\ref{tab:physical_properties}).  Assuming similar beaming fractions, this makes them close to an order of magnitude less energetic compared to the weakest bursts seen from FRB~121102 to date\cite{gourdji2019}, and between four to six orders-of-magnitude less energetic compared to FRB~180924\cite{bannister2019} and FRB~190523\cite{ravi2019}.  Unless multiple models are invoked, a viable model for FRBs must address this large range of (apparent) energy outputs.  

Comparing instead with pulsar emission, we note that the bursts from \frb\ are still at least a million times more energetic compared to the brightest giant pulses seen from the Crab pulsar --- suggesting that they are not simply an exceptionally bright version of the known pulsar giant pulse phenomenon\cite{lyutikov2017}.  

As previously suggested\cite{chime2019c}, the proximity of \frb is an advantage for multi-wavelength follow-up of the host galaxy and local environment.  Whereas targeted observations of FRB~121102 have failed to detect either prompt or persistent optical, X-ray or gamma-ray counterparts\cite{scholz2016,scholz2017,hardy2017,magic2018}, similar observations towards \frb\ may strongly constrain magnetar-based models, even in the event of non-detections. \frb is thus one of the most promising known sources for understanding the nature of FRBs.

\clearpage

\clearpage

\clearpage

\clearpage

\begin{table}
\centering
    \caption{\label{tab:physical_properties}
    {\bf Burst properties.}
    Widths, fluences and spectral energy densities are determined using the Effelsberg auto-correlation data of the bursts from \frb, with a centre frequency of 1.7~GHz and dedispersed for a DM~$= 348.76\ \mathrm{pc\ cm^{-3}}$. W$_{\text{tot}}$ is the FWHM of the total burst envelope, and W$_{\text{sub}n}$ represents the FWHM of any sub-bursts, ordered according to their arrival time (earliest sub-burst is labelled ``sub1''). The coloured boxes correspond to the bars in Figure~\ref{fig:efauto_profile} highlighting which structure is associated with the quoted widths. The fluence is determined by integrating the burst profile over W$_{\text{tot}}$ and is converted to physical units using the radiometer equation\cite{cordes2003} (see Methods). The spectral energy density is estimated using the measured fluence and luminosity distance.
    The position offsets ($\Delta\alpha, \Delta\delta$) are referred to the average J2000 burst position of $\alpha = 01^\text{h}58^\text{m}00.7502^\text{s},\ \delta = +65^\circ 43^\prime 00.3152^{\prime\prime}$. The flux densities, $S_\nu$, are measured by fitting a circular Gaussian to the EVN visibility data and are averages over the gate width. 
    \newline $^{\rm a}$ At 1.7~GHz reference frequency, and corresponding to the sum of all sub-bursts.  A conservative fractional error of 30\% is adopted for the derived fluences and energy densities.  These are assumed to be isotropic.\newline
    $^{\rm b}$ At 1.7~GHz reference frequency.  The absolute flux scale may exhibit an additional $\sim $15\% uncertainty due to possible systematic gain calibration offsets.\newline
    $^{\rm c}$ See Methods for how the total burst width of B4 was determined.}
\medskip
\resizebox{\textwidth}{!}{
\begin{tabular}{cccccccccc}
\hline
Burst &  W$_{\text{tot}}$ \fcolorbox{black}{cyan}{\rule{0pt}{4pt}\rule{4pt}{0pt}}\quad & W$_{\text{sub1}}$ \fcolorbox{black}{orange}{\rule{0pt}{4pt}\rule{4pt}{0pt}}\quad & W$_{\text{sub2}}$ \fcolorbox{black}{purple}{\rule{0pt}{4pt}\rule{4pt}{0pt}}\quad
& W$_{\text{sub3}}$ \fcolorbox{black}{green}{\rule{0pt}{4pt}\rule{4pt}{0pt}}\quad & Fluence & Spectral energy density & $\Delta\alpha$ & $\Delta\delta$ & $S_\nu$ \\
& (ms) & (ms) & (ms) & (ms) & (Jy~ms)$^\text{a}$ & (10$^\text{28}$~erg~Hz$^{-1}$)$^\text{a}$ & (mas) & (mas) & (Jy)$^\text{b}$\\
\hline
B1 & 1.86 $\pm$ 0.13 & -- & -- & -- & 0.72  & 1.90 & $-$2.1 $\pm$ 3.2 & \phantom{$-$}1.1 $\pm$ 3.2 & 0.74 $\pm$ 0.08\\
B2 & 0.24 $\pm$ 0.02 & -- & -- & -- & 0.20 & 0.53 & $-$1.5 $\pm$ 2.9 & $-$0.2 $\pm$ 2.9 & 0.66 $\pm$ 0.05\\
B3 & 1.72 $\pm$ 0.14 & 0.14 $\pm$ 0.02 & 0.12 $\pm$ 0.01 & --   & 0.62 & 1.64 & \phantom{$-$}2.8 $\pm$ 2.8 & $-$1.4 $\pm$ 2.8 & 0.50 $\pm$ 0.04\\
B4 & 1.66 $\pm$ 0.05$^{\rm c}$  & 0.24 $\pm$ 0.01 &  0.06 $\pm$ 0.0006 & 0.68 $\pm$ 0.03  & 2.53 & 6.68 & \phantom{$-$}0.1 $\pm$ 1.9& \phantom{$-$}0.3 $\pm$ 1.9 & 2.3 $\pm$ 0.3 \\
\hline
\end{tabular}}
\end{table}

\clearpage

\newpage

\begin{methods}

\subsection{A priori localisation.}

The CHIME/FRB Collaboration discovered\cite{chime2019c} the repeating source \frb and refined its position using the CHIME/FRB baseband mode. The automatic baseband triggering system of CHIME/FRB\cite{chime2018} captured\cite{chime2019c} raw voltages for a burst from \frb.  The intensity of the signal in the region surrounding the original detection was then calculated by producing a grid of tightly-packed tied-array beams.  A refined position was obtained by fitting this intensity map with a simple model of the telescope beam.  The technique will be described in more detail in a future paper (Michilli et al. in prep.).  Since the analysis strategy was still in a preliminary stage at the time, we estimate a systematic uncertainty on this position of roughly three arcminutes, based on a small sample of four known pulsars.  CHIME/FRB baseband localisations are expected to have higher precision in the future.

\subsection{European VLBI Network and Effelsberg observations.}

The EVN observed the field of \frb on June 19th, 2019 for 5.5~h at a central frequency of 1.7~GHz and with a bandwidth of 128~MHz.  The phase centre was placed at the position provided by the initial CHIME/FRB baseband localisation: $\alpha = 01^\text{h}57^\text{m}43.2^\text{s},\ \delta = 65^\circ 42^{\prime}01.02^{\prime\prime}$ (J2000 coordinates).

A total of eight dishes participated in the EVN observations: the 100-m Effelsberg, the 65-m Tianma, the 32-m Medicina, the 32-m Toru\'{n}, the 32-m Irbene, the 25-m $\times$ 38-m Jodrell Bank Mark II, the 25-m Onsala, and a single 25-m dish from the Westerbork array.
The data were streamed in real time (e-EVN setup) to the EVN Software Correlator (SFXC\cite{keimpema2015}) at the Joint Institute for VLBI ERIC (JIVE) in Dwingeloo, The Netherlands. The real-time visibility data are comprised of eight 16-MHz subbands of 32 channels each, with full circular polarisation products, and 2-s time averaging.

In parallel, we buffered the individual station raw voltage data in order to allow high-time-resolution correlations afterwards, at the times of any detected bursts. This method allows one to recover the signal from any position within the primary beam of the antennas (which have a full-width at half-maximum, FWHM, of roughly 7~arcmin in the case of the 100-m Effelsberg dish, and roughly 30~arcmin for the 25-m antennas), which would otherwise be smeared due to time and frequency averaging if it is more than tens of arcseconds away from the phase centre used in cross-correlation.
We observed 3C454.3 and J0745+1011 as fringe-finders and bandpass calibrators. J0207+6246 was observed as phase calibrator (located 3.1$^\circ$ away from \frb) in a phase-referencing cycle of 2~min on the calibrator and 5~min on the target, \frb. We also observed J0140+6346 as a check source (3.1$^\circ$ away from the same phase calibrator) following the same phase-referencing cycle.

Simultaneously, we recorded high-time-resolution filterbank data at 1.7~GHz using the 100-m Effelsberg telescope and the PSRIX data recorder\cite{lazarus2016}, which is designed for pulsar observations. We observed \frb for a total on-source time of 3.47~h in the frequency range 1597--1737~MHz. This total bandwidth was divided into 144 spectral channels of 0.98~MHz each. Of these, the bottom 48 channels were corrupted by radio frequency interference (RFI), likely from Iridium satellites. Consequently, these channels were removed from the data prior to beginning the analysis, giving a usable bandwidth of 93.75~MHz.  The data were recorded with full Stokes information and a time resolution of 81.92~$\upmu$s. Before the beginning of the observation, we performed a short test scan on the known pulsar PSR~B2111+46, in order to verify the data integrity.

\subsection{Interferometric data reduction.}

The EVN data were analysed using standard procedures within the Astronomical Image Processing System ({\tt AIPS}\cite{greisen2003}) and {\tt Difmap}\cite{shepherd1994} software packages. {\it A priori} amplitude calibration was performed using the known gain curves and system temperature measurements recorded at each station during the observations.
Poor data, primarily due to RFI, were flagged manually ($\lesssim 10$\% of the total data). The remaining data were then fringe-fitted and bandpass calibrated using the fringe-finders and phase calibrator, which were imaged and self-calibrated to improve the final calibration of the data. The obtained solutions were transferred to \frb and J0140+6346 before creating the final images.

The check source J0140+6346 was used to estimate both the absolute astrometric uncertainty and the potential amplitude losses that could have been introduced due to the phase-referencing technique. Although the latter accounted for less than approximately 10\% of the total flux density scale, we found a significant offset ($\sim 4$~mas) of the centroid of the check source position with respect to its known coordinates from the International Celestial Reference Frame (ICRF). The origin of this offset is explained by both the uncertainties associated with the phase-referencing technique\cite{chatterjee2004,pradel2006,kirsten2015} and the extended structure of J0207+6246, the phase calibrator. The core of this source shows multiple components that make the determination of the true position of the source ambiguous at the milli-arcsecond level. We corrected for this observed offset in all burst positions presented in this study, but an uncertainty on the final absolute positions of $\pm 1.7\ \text{mas}$ and $\pm 2.1\ \text{mas}$ in right ascension and declination, respectively, still remains due to the ambiguity in the exact position of this reference source and its extension (the source is resolved on milliarcsecond scales).

\subsection{Search for \frb bursts.}

The PSRIX data were analysed in order to search for dispersed, millisecond-duration bursts, using the {\tt PRESTO}\cite{ransom2001} software package. The tool {\tt rfifind} was used to identify frequency channels and time samples contaminated by RFI. We estimate that $\sim 4\%$ of the data were masked using {\tt rfifind}.  Incorporating the RFI mask, the data were then incoherently dedispersed to DMs in the range 0--700~pc~cm$^{-3}$ in steps of 1~pc~cm$^{-3}$ using {\tt prepsubband}, generating dedispersed time series at each trial DM. To search for short-duration bursts, each dedispersed time series was convolved with boxcar functions of various widths, using {\tt single\_pulse\_search.py}. Given the frequency resolution of the PSRIX data, the residual dispersive delay within each channel results in a temporal smearing of $\sim 0.6$~ms at the expected\cite{chime2019c} DM of \frb (349~pc~cm$^{-3}$). This intra-channel smearing dominates over the sampling time, causing a loss of sensitivity to extremely narrow bursts. We were sensitive to bursts with widths exceeding the intra-channel smearing time, and up to 98.3~ms. A DM-time plot (known as a single-pulse plot, generated by {\tt single\_pulse\_search.py}) was created for each scan and inspected by eye for candidates. For each candidate identified, the dynamic spectrum was generated and inspected by eye to distinguish between astrophysical signals and terrestrial sources of RFI. During our 3.47-h on-source time, we found four bursts from \frb at its known DM. As a final check, each dedispersed time series was filtered for RFI in the search for candidates with peak signal-to-noise ratio S/N $> 7$, using an automated classifier\cite{michilli2018b,michilli2018c} that returned only the same four bursts found in the initial manual search.

Using the time of arrival of each individual burst, one-second duration Effelsberg auto-correlations were created for each burst, using the recorded EVN raw-voltage data. The auto-correlation data were generated with a higher temporal resolution (16-$\upmu$s time bins) and frequency resolution (62.5-kHz channels) compared with the PSRIX data, and were coherently dedispersed to 350~pc~cm$^{-3}$. Coherent dedispersion mitigates the effects of intra-channel smearing to less than 3 $\upmu$s for DMs that are within a few units of 350~pc~cm$^{-3}$. This provided us with the opportunity to better study the burst structure (limited only by our temporal resolution and the signal-to-noise), which is unresolved in the PSRIX data.

We identified the four bursts from \frb in the auto-correlations (Figure \ref{fig:efauto_profile} shows their band-averaged burst profiles and dynamic spectra). For comparison, we also show the profiles and dynamic spectra for the bursts in the PSRIX data (Extended Data Figure \ref{fig:efpsr_profile}).

\subsection{Burst properties.}

    Using the brightest burst, labelled B4, we determined a precise DM for \frb. We used the {\tt PSRCHIVE}\cite{hotan2004} tool {\tt pdmp} to determine the DM that corresponds to a peak in the signal-to-noise. Because B4 has a bright and narrow sub-burst, optimising the signal-to-noise is likely also equivalent to optimising the structure of the burst\cite{hessels2019}, in this case.  We searched DMs of 350 $\pm$ 3~pc~cm$^{-3}$ in steps of 0.01~pc~cm$^{-3}$ using {\tt pdmp}, and found the optimal DM to be 348.76 $\pm$ 0.10~pc~cm$^{-3}$ (consistent with the DM of 349.2 $\pm$ 0.3~pc~cm$^{-3}$ determined for this source by CHIME/FRB\cite{chime2019c}). The bursts shown in both Figure~\ref{fig:efauto_profile} and Extended Data Figure~\ref{fig:efpsr_profile} are dedispersed to this DM value.

Table~\ref{tab:physical_properties} lists the physical properties of the bursts using the Effelsberg auto-correlations, dedispersed to 348.76~pc~cm$^{-3}$. We measured the burst FWHM duration and peak time using a Gaussian fit\cite{hessels2019}. The brightest burst, B4, shows a complicated structure (as can be seen in Figure \ref{fig:efauto_profile}). The width of each individual sub-burst in B4 was determined using the same Gaussian fit as above. The total width of B4, however, was measured by defining the beginning of the burst as the peak time of the first sub-burst minus W$_{\text{sub1}}$, and the end of the burst as the peak time of the last sub-burst plus W$_{\text{sub3}}$.

To determine the fluence, we define a larger time window, determined by eye, than the quoted FWHM durations, which encompasses all of the burst flux. First, we produce a time series containing each burst by summing over all frequency channels. We define an off-pulse region, and use this to normalise the time series. We then convert the signal in each time bin to physical units by using the radiometer equation\cite{cordes2003}, and sum over the burst duration to estimate the fluence. It should be noted that B1 and B3 were downsampled by a factor of four for this analysis, resulting in a lower time resolution than the other bursts. For the radiometer equation, we take typical values for the system temperature $T_{\text{sys}} \approx 20\ \text{K}$ and telescope gain $G \approx 1.54\ \text{K\ Jy}^{-1}$ for Effelsberg. For the fluence values quoted in Table~\ref{fig:efauto_profile}, we provide a conservative fractional error of 30\%. Also shown in Table~\ref{tab:physical_properties} are the burst spectral energy densities\cite{law2017}, using the derived burst fluences and luminosity distance to \frb.  All four bursts are significantly dimmer compared to those previously detected\cite{chime2019c} with CHIME/FRB, which is unsurprising given the higher sensitivity of the 100-m Effelsberg telescope.

Extended Data Table~\ref{tab:detection_properties} shows the detection properties of the bursts, as seen in both the PSRIX data and the Effelsberg auto-correlations. For the PSRIX data, we quote the maximum S/N detected using {\tt PRESTO}'s {\tt single\_pulse\_search.py} with an associated DM. The burst S/N for the auto-correlations is that measured at DM $= 348.76$~pc~cm$^{-3}$, again using {\tt single\_pulse\_search.py}. The burst width in Extended Data Table~\ref{tab:detection_properties} is the boxcar width used in {\tt single\_pulse\_search.py} resulting in the peak S/N value. The S/N of the four bursts, as discovered in the PSRIX data, are shown in Extended Data Figure~\ref{fig:scans} as a function of time during our observation. The first three bursts appear clustered in time, occurring within 1.5~h from the beginning of our observation.

To study frequency-dependent brightness variations in the dominant sub-burst of B4 --- which are expected if there is scintillation --- we conducted a standard auto-correlation analysis\cite{cordes1985}. We generated the spectrum of the burst within the FWHM duration (W$_{\rm sub2}$ = 0.06 ms) using Effelsberg auto-correlations with a frequency resolution of 15.6~kHz. Spectral channels containing significant RFI, along with those at the edges of the 16-MHz subbands, were set to zero in the spectrum before generating the auto-correlation function (ACF).

We calculated the ACF using the following expression:
\begin{equation}\label{eq:autocorr2}
    {\rm ACF}(\Delta \nu) = \frac{\sum\limits_{i}(S(\nu_{i})-\bar{S})(S(\nu_{i}+\Delta\nu)-\bar{S})}{\sqrt{\sum\limits_{i}(S(\nu_{i})-\bar{S})^2}\sqrt{\sum\limits_{i}{(S(\nu_{i}+\Delta\nu)-\bar{S})^2}}},
\end{equation}
where $\nu$ is the frequency, $\Delta\nu$ is the frequency lag, and $\bar{S}$ is the mean of the spectrum. In this expression each summation is over the indices $i$ which give non-zero values for both $S(\nu_{i})$ and $S(\nu_{i}+\Delta\nu)$ (i.e. we only include the ``good" data in the auto-correlation analysis). The ACF calculated using Equation \ref{eq:autocorr2} is shown in Extended Data Figure \ref{fig:ACF}.

The central part of the ACF (for lags in the range $-$1.016 to $+$1.016 MHz) after removing the zero lag was then fitted with a Lorentzian function\cite{rickett1990} of the form:
\begin{equation}
    \dfrac{a}{x^2+{\Delta\nu_{d}}^2}+b,
\end{equation}
using a least squares fit, for free parameters $a$, the amplitude; $\Delta\nu_{d}$, the decorrelation bandwidth (also known as the scintillation bandwidth: defined\cite{cordes1985} as the half-width at half maximum of the ACF); and offset $b$.  Extended Data Figure \ref{fig:ACF} shows the fit to the ACF. The Lorentzian fit, with 127 degrees of freedom, returned a $\chi^2$ of 105.2. We assume that the ACF between 3 and 20~MHz lags contain no structure and represents the noise in the central profile of the ACF. We measure the standard deviation in this region and use this as our uncertainty for the calculation of $\chi^2$. The $p$-value of the fit is 0.92.

We measure the scintillation bandwidth to be 59 $\pm$ 13~kHz. This estimate is consistent with the NE2001 model\cite{cordes2002} prediction of $\sim$ 60~kHz along the line-of-sight at our observing frequency of 1.7~GHz. Using the simple relationship
\begin{equation}
    2\pi\tau\Delta\nu_{d}\sim 1,
\end{equation}
we estimate a scattering timescale, $\tau$, of $\sim$ 2.7 $\upmu$s, which we find is also similar to the NE2001 model prediction of $\sim$ 2 $\upmu$s along the line-of-sight\cite{cordes2002}.  This constraint on the scattering time is much tighter compared to CHIME/FRB measurements\cite{chime2019c} of \frb bursts, even after accounting for the different observing frequencies and the typical $\nu^{-4}$ scaling of scattering time.  CHIME/FRB nominally measured scattering times of approximately 2~ms for two \frb bursts, but these measurements are likely a reflection of burst morphology as opposed to genuine scattering effects (as was previously considered\cite{chime2019c}).

\subsection{High-time-resolution correlation and burst imaging.}

Coherently dedispersed visibilities for each burst were created by re-correlating the EVN data for only the time ranges corresponding to the burst durations. The exact time ranges were initially determined from the Effelsberg PSRIX data and then refined using coherently dedispersed Effelsberg auto-correlations. From the resulting auto-correlations we created pulse profiles and used these to set the optimal correlator gates applied to the interferometric data.

The {\it a priori} position of \frb provided at the time of the EVN observation had an uncertainty of $\sim 3$~arcmin (see above), making it infeasible to localise the bursts by directly imaging this region at milli-arcsecond resolution (which would require an image with $\sim 10^{11}$ pixels). However, for strong signals, it is possible to estimate the position of the source from the geometric delays across the different telescope baselines. The differential geometric phase delay, $\Delta\phi_g$, of the source with respect to the phase centre ({\it a priori} pointing position) for a given baseline is given by\cite{fomalont1999,thompson1999}:
\begin{equation}
    \Delta\phi_g = 2 \pi \nu ( u\,\Delta\alpha\, \cos\delta + v \Delta\delta)
\end{equation}
where $\nu$ is the observed frequency, $(\Delta\alpha, \Delta\delta)$ is the positional offset of the source with respect to the phase centre in right ascension ($\alpha$) and declination ($\delta$), respectively, and $(u,v)$ are the coordinate offsets for the given baseline.

We used the strongest burst, B4, to estimate the offset of the source with respect to the phase centre. We then re-correlated the data at this refined J2000 position of $\alpha = 01^\text{h}58^\text{m}00.5^\text{s},\ \delta = 65^\circ 43^\prime 01.0^{\prime\prime}$ (with an estimated uncertainty of a few arcseconds), which was $\sim 2\ \text{arcmin}$ from the {\it a priori} position --- and thus consistent within the estimated uncertainties.  We applied the previously described EVN calibration to these burst data and imaged the individual bursts. All bursts were detected above a $7$-$\sigma$ confidence level. Different imaging schemes were used during the imaging process in order to emphasise resolution (i.e. contribution from the longest baselines) or sensitivity (i.e. contribution from the core of the array). The final images were produced by using a natural weighting and excluding data from the Tianma station which, despite providing the highest resolution, exhibited potential calibration issues that could affect the absolute burst positions by a few milli-arcseconds --- especially for such short integration times, where the $uv$ coverage is poor. Table~\ref{tab:physical_properties} lists the burst properties obtained from the EVN images, which are shown in Figure~\ref{fig:evn-bursts}. An image of the combined burst data was also produced. This final image is dominated by the emission of the strongest burst, B4, and thus we only consider the results from the four bursts individually.

The flux densities and positions were measured using {\tt Difmap} and the Common Astronomy Software Applications ({\tt CASA}) software packages by fitting a circular Gaussian distribution component to the detected emission in the $uv$ plane, which is expected to be a robust method against station gain calibration uncertainties\cite{natarajan2017}.
The position of each burst was measured independently and then we determined a weighted average J2000 position for the source of $\alpha = 01^\text{h}58^\text{m}00.7502^\text{s} \pm 2.3\ \text{mas},\ \delta = 65^\circ 43^\prime 00.3152^{\prime\prime} \pm 2.3\ \text{mas}$, where we used the fluence of each burst as weights.  This is within 1.7~arcsec of the position determined via mapping the geometric delays.  The final positional uncertainty reflects the statistical uncertainties from the individual position measurements ($\sim$ 1~mas), the uncertainties in the absolute ICRF position of the phase calibrator (J0207+6246; 0.15~mas) and check source (J0140+6346; 0.1~mas), and the systematic uncertainty associated with the phase-referencing technique and the extended structure of the check source and phase calibrator ($\sim$2~mas), as explained previously. Uncertainties were added in quadrature.

Finally, we searched for continuum radio counterparts of \frb by imaging the full interferometric data set with a field-of-view of 2$\times$2~arcsec$^{2}$ centred at the position of the bursts (and thus also covering the full extent of the nearby star-forming region). We detected no emission at the position of \frb as well as no significant ($< 4.5\sigma$) compact emission anywhere in the field above the r.m.s. noise level of $10\ \upmu\text{Jy\ beam}^{-1}$ ($19\ \upmu\text{Jy\ beam}^{-1}$ when excluding the data from the Tianma station). We also searched for possible emission from the core of the galaxy, but found no signal above the 6-$\sigma$ r.m.s. noise level in a 4$\times$4~arcsec$^{2}$ area.

\subsection{VLA observations.}

In June 2019, we used the Karl G.\ Jansky Very Large Array (VLA) to perform a quasi-simultaneous search for millisecond transients and faint persistent emission associated with \frb. The VLA observed from 1--2~GHz with baseline lengths up to 11~km, giving a synthesised beam size of roughly 4~arcsec. VLA visibilities were sampled at 5-ms time resolution and searched in real time by the Realfast system\cite{law2018}; the same data were integrated at 3-s time resolution and calibrated with the standard {\tt CASA} calibration pipeline. In a total of 14~h of observing (with on-target efficiency of 80\%), we found no millisecond transients to an 8-$\sigma$ limit of approximately 50~mJy~beam$^{-1}$ in 5~ms or, equivalently, a fluence limit of 0.25~Jy~ms. No VLA observing was coincident with known burst times detected by the EVN. We concatenated all 14~h of data, applied an iteration of phase self-calibration using relatively bright background sources in the field of view of the antennas, and imaged 600~MHz of the bandwidth centred near 1.6~GHz to search for persistent radio emission on arcsecond scales. We confirmed that the astrometric accuracy of the deep image is good to better than 1~arcsec by associating seven bright field radio sources to the catalogued location in the NRAO VLA Sky Survey (NVSS\cite{condon1998}). No persistent emission is detected at the position of \frb brighter than a 3-$\sigma$ limit of 18~$\upmu$Jy\ beam$^{-1}$.  A 25-$\upmu$Jy source is located at a nearby J2000 position of $\alpha = 01^\text{h}58^\text{m}00.9^\text{s},\ \delta = 65^\circ43^\prime07^{\prime\prime}$ (see Extended Data Figure~\ref{fig:vla-map}), but this position is offset by approximately 7~arcsec from the location of \frb, which is much more than the astrometric precision, estimated to be 0.4~arcsec (the VLA observed in its B configuration at the time).  It is thus very unlikely to be associated with \frb or the host galaxy.

\subsection{Optical observations and host galaxy.}

The EVN localisation of \frb places it at the outskirts of SDSS~J015800.28+654253.0, an apparent Sb spiral galaxy (possibly with a faint bar) with an estimated\cite{alam2015} photometric redshift $z_\text{phot}=0.07 \pm 0.05$.

We obtained imaging and spectroscopic observations of the host galaxy using the GMOS spectrograph on the Gemini-North telescope between July and September 2019 (Program ID GN-2019A-DD-110). We acquired 2$\times$900-s exposures with each of the $g^\prime$ and $r^\prime$ filters. The images were processed with the {\tt gemini IRAF/pyraf} package, and sources were extracted using {\tt Source Extractor}\cite{bertin1996}. The images were de-warped and astrometrically corrected by matching the point source positions to those from the \emph{Gaia} Data Release 2 (DR2) Catalogue\cite{gaia2016,gaia2018}. The matching precision was 17--20~mas r.m.s. for each image. The seeing was measured to be 0.99 and 0.78~arcsec in the $g^\prime$ and $r^\prime$ images, respectively.

Figure~\ref{fig:galaxy} shows the $r^\prime$ image of the galaxy, along with the position of \frb. The galaxy is nearly face-on, with spiral arm structures visible.  \frb appears co-located with one of the outer arms of the galaxy, near a bright clump in the $r^\prime$ image (see Extended Data Figure~\ref{fig:galaxy-zoom} for a more detailed zoom on this region). We attribute this emission to that of \Halpha, which is bright in emission in the GMOS spectra and, at the redshift of the host galaxy, falls within the $r^\prime$ image.  The total stellar mass of the galaxy has been estimated to be $\sim 10^{10}$ times the mass of the Sun by using the WISE W1 and W2 colours\cite{jarrett2017}, which provide a more robust estimation given that they are less affected by extinction.

We performed long-slit optical spectroscopy, simultaneously targeting the galaxy core and the FRB location, in order to measure the redshift and spectral properties of the galaxy and the location of \frb. We acquired 4$\times$900-s spectroscopic observations with a 1.5~arcsec slit and the R400 grating --- along with corresponding flat fields, copper-argon arcs, and observations of BD\,+28\,4211 to correct for sensitivity and extinction.  As shown in Extended Data Figure~\ref{fig:galaxy-zoom}, the slit was rotated to align with the host galaxy centre and FRB position.

The spectroscopic data were processed with the {\tt Gemini IRAF/pyraf} package. The FWHM of the emission lines from the sky was measured to be 7--8~\AA, matching the theoretical resolving power of $R\approx 640$. The wavelength calibration with copper-argon emission lines had a precision of 0.7--0.8~\AA. However, the lack of arc lines towards the bluer edge reduces the accuracy of the wavelength solution towards 5200~\AA.

We extracted the spectrum within 5~arcsec of the galaxy centre as well as 2~arcsec around the location of \frb itself. We used the trace of the galaxy centre, offset by 7.7~arcsec to trace the location of \frb. Figure~\ref{fig:galaxy_spectra} shows the calibrated spectra at each location separately. We identify redshifted \Halpha, $\mathrm{N_{II}}$, $\mathrm{S_{II}}$ and $\mathrm{O_{III}}$ lines with an average redshift of $z = 0.033 \pm 0.001$ if all lines are included and $z = 0.0337 \pm 0.0002$ when excluding the $\mathrm{O_{III}}$ lines that are at the blue edge of the CCD (and hence may have reduced wavelength calibration accuracy). Extended Data Table~\ref{tab:spectral_lines} shows the line centroids, widths and fluxes. The spectrum at the location of \frb is dominated by emission lines with little continuum emission. The $\mathrm{O_{III}}$ 5007\AA\ line is detected at the FRB location but was corrupted by a cosmic-ray hit. As the slit was positioned at the edge of the nearby star-forming clump, we observe a non-uniform slit illumination pattern and the line centroids, especially for \Halpha, are shifted towards the redder wavelengths by $\sim 1$~\AA. Nevertheless, we can confirm that both the clump and the galaxy are at the same distance.

The obtained flux for the \Halpha emission at the position of \frb (which is dominated by the star-forming clump) implies a luminosity of $L_{\text{H}\alpha} \sim (2.0 \pm 0.1) \times 10^{39}$~erg~s$^{-1}$,  which corresponds to a star formation rate\cite{kennicutt1994} of
$\approx 0.016\ \text{M}_\odot\ \text{yr}^{-1}$.  These values were extracted from a region of $1.5 \times 2$~arcsec$^2$ around the FRB, corresponding to 1.5\,kpc$^2$, implying a star-formation surface density of $\approx 10^{-2}~\mathrm{M_\odot~yr^{-1}~kpc^{-2}}$.
Note that these values can be significantly affected by extinction, which is estimated to be $E(B-V) \approx 1.01$.

From the \Halpha, $\mathrm{N_{II}}$, and $\mathrm{S_{II}}$ line fluxes, we estimate the metallicity of the galaxy\cite{dopita2016} as $12 + \log(\mathrm{O}/\mathrm{H}) = 8.82$, close to solar neighborhood metallicity, and five times higher when compared to the host of FRB~121102 ($<8.1$ on the same metallicity scale\cite{tendulkar2017}).

\subsection{Chance alignment probability with host galaxy.}

The projected position of \frb is coincident with a star-forming clump in the spiral galaxy SDSS~J015800.28+654253.0. To confirm an association with the galaxy we have estimated the probability of chance coincidence between the two sources.

We have considered a uniform density of galaxies across the sky. Assuming a Poisson distribution of galaxies in the relevant co-moving volume, we estimate the number density of typical galaxies brighter than $M_B = -16$ (i.e., galaxies with masses greater than $\sim$ 40\% of the dwarf galaxy host\cite{tendulkar2017} of FRB~121102) based on the all-galaxy luminosity function\cite{faber2007,blanton2003}, and their typical size distribution based on the galaxies reported in SDSS~DR7\cite{zhang2019}. Then, we calculate the chance coincidence probability for a source to be co-located within twice the median half-light radius of any galaxy as a function of redshift.

At the measured redshift of $z \sim 0.0337$, the chance coincidence probability $P \ll 0.1\%$. However, we may consider the probability up to a redshift of $\sim 0.11$, the maximum redshift considered for \frb based on the observed DM\cite{chime2019c}. Even in this case $P \lesssim 1\%$ (see Extended Data Figure~\ref{fig:probability}). The association between \frb and the galaxy is thus statistically robust.

\subsection{Dispersion measure budget}

\frb is close to the Galactic plane, with longitude $l = 129.7^{\circ}$ and latitude $b = 3.7^{\circ}$.  The total measured DM of \frb is DM$_{\rm Tot} =$ \dm, and is expected to be the sum of contributions from the Milky Way disk (DM$_{\rm MW}$) and halo (DM$_{\rm Halo}$), the intergalactic medium (DM$_{\rm IGM}$), and host galaxy (DM$_{\rm Host}$; including both the ISM and halo of the host).  For the line-of-sight to \frb, DM$_{\rm MW}$ is predicted to be 199~\dmu and 325~\dmu by the NE2001 and YMW16 Galactic electron density models, respectively\cite{cordes2002,yao2017}.  DM$_{\rm Halo}$ is estimated to be 50--80~\dmu or approximately $45$~\dmu according to two available models\cite{prochaska2019,yamasaki2019}.  Given the range of model predictions, one can ascribe up to roughly 100~\dmu to DM$_{\rm IGM}$ and DM$_{\rm Host}$.  For redshift $z = 0.0337$, DM$_{\rm IGM}$ is expected to be\cite{inoue2004} approximately 34~\dmu --- though this estimate is highly line-of-sight dependent for low redshift\cite{li2019}.  Overall, after subtracting these various contributions, DM$_{\rm Host}$ is likely less than $70$~\dmu, and could be much smaller.  The YMW16 model over-predicts DM$_{\rm MW}$ in this direction (unless the DM$_{\rm Halo}$, DM$_{\rm IGM}$ and DM$_{\rm Host}$ contributions are all much smaller than expected), and this illustrates that FRB DMs will help constrain Galactic electron density models.  Conversely, because the line-of-sight to \frb passes through the Galactic plane, with a large but uncertain Galactic DM contribution, it is unlikely to be a good probe for refining the DM-z relationship in the IGM\cite{inoue2004}.

\subsection{\frb compared with the other localised repeating fast radio burst source, FRB~121102.}

The physical origin of the FRB phenomenon remains unclear, and many of the proposed models\cite{platts2018} remain consistent with the observational facts we present here for \frb.  We now briefly consider the nature of \frb by comparing to two of the families of models proposed for FRB~121102, which is the FRB source studied in most detail to date, and the only other well-localised repeater.  

One of the models to explain the associated persistent radio counterpart to FRB~121102 proposes the existence of a 20--50 year-old supernova remnant and nebula powered by a young flaring magnetar\cite{lyubarsky2014,beloborodov2017,margalit2018,metzger2019}. The absence of a similarly bright radio nebula associated with \frb could be explained, in this model, as an older system whose emission has already faded. Assuming a similar model\cite{margalit2018}, the upper limits on persistent radio emission associated with \frb are consistent with a system that is at least about five times older than FRB~121102, and thus $\gtrsim 200-500$~yr old. Additionally, the same model also predicts a characteristic decay of the RM with time. An RM as large as the one observed towards FRB~121102\cite{michilli2018} would drop to the RM observed\cite{chime2019c} towards \frb at a source age of $\sim 300$~yr.  However, while the results we present here are conceivably consistent with the young magnetar scenario for repeating FRBs, we find no new support for this scenario.

Other models, inspired originally by FRB~121102, invoke interaction between a nearby strong plasma stream and a neutron's magnetosphere\cite{zhang2017}.  In the case of FRB~121102, the large\cite{michilli2018} RM and the observational similarity\cite{marcote2017} of its compact, persistent radio source to low-luminosity active galactic nuclei (LLAGN) suggest that the bursts could originate from a massive black hole, or a neutron star in the near vicinity.  An accreting massive black hole can provide a plasma stream that could interact with a neutron star magnetosphere\cite{zhang2018}.  The absence of a persistent radio counterpart to \frb, and its location in the spiral arm of a morphologically well-defined galaxy, make such a model arguably less likely in the case of \frb.  For a black hole with a similar mass to that of the one considered for FRB~121102 ($10^{5\text{--}6}$~M$_{\odot}$), our upper limits would imply a very low accretion rate of $\sim 10^{-7}\ L_\text{Edd}$. Nonetheless, this accretion rate is still compatible with the low end observed in LLAGN\cite{loewenstein2001}. The presence of a massive black hole at the apex of the star-forming clump is thus not excluded, although rather unlikely.  Nonetheless, in an interacting model\cite{zhang2017}, the plasma stream can also have a different origin, and it is conceivable that \frb\ is a neutron star interacting with, e.g., the jet of a stellar-mass black hole.

The proximity of \frb\ will aid in performing deep multi-wavelength observations, which may help to discriminate between the various scenarios outlined above, while potentially also suggesting new ones.

\end{methods}
\begin{addendum}

\item[Data Availability] The datasets generated from the EVN observations and analyzed in this study are available at the Public EVN Data Archive (\url{http://www.jive.eu/select-experiment}) under the experiment code EM135C. The datasets generated from the Gemini observations are available at the Public Gemini Observatory Archive (\url{https://archive.gemini.edu}) with Program ID GN-2019A-DD-110.
    
    \item[Code Availability] The codes used to analyse the data are available at the following sites:
    AIPS (\url{http://www.aips.nrao.edu/index.shtml}),
    CASA (\url{https://casa.nrao.edu}),
    Difmap (\url{ftp://ftp.astro.caltech.edu/pub/difmap/difmap.html}),
    IRAF (\url{http://ast.noao.edu/data/software}),
    PRESTO (\url{https://github.com/scottransom/presto}),
    and PSRCHIVE (\url{http://psrchive.sourceforge.net}).
    
    \item
    We thank W.~J.~G.~de~Blok, L.~Connor, N.~Maddox, E.~Petroff, H.~Vedantham and J.~Weisberg for valuable discussions.
    The European VLBI Network is a joint facility of independent European, African, Asian, and North American radio astronomy institutes. Scientific results from data presented in this publication are derived from the following EVN project code: EM135.
    This work was also based on simultaneous EVN and PSRIX data recording observations with the 100-m telescope of the MPIfR (Max-Planck-Institut f\"{u}r Radioastronomie) at Effelsberg, and we thank the local staff for this arrangement.
    Our work is also based on observations obtained at the Gemini Observatory (program DT-2019A-135), which is operated by the Association of Universities for Research in Astronomy, Inc., under a cooperative agreement with the NSF on behalf of the Gemini partnership: the National Science Foundation (United States), the National Research Council (Canada), CONICYT (Chile),  Ministerio de Ciencia, Tecnolog\'{i}a  e Innovaci\'{o}n Productiva (Argentina), and Minist\'{e}rio da Ci\^{e}ncia, Tecnologia e Inova\c{c}\~{a}o (Brazil).
    B.M. acknowledges support from the Spanish Ministerio de Econom\'ia y Competitividad (MINECO) under grants AYA2016-76012-C3-1-P and MDM-2014-0369 of ICCUB (Unidad de Excelencia ``Mar\'ia de Maeztu'').  J.W.T.H. acknowledges funding from an NWO Vidi fellowship and from the European Research Council under the European Union's Seventh Framework Programme (FP/2007-2013) / ERC Starting Grant agreement nr. 337062 (``DRAGNET'').  M.B. is supported by an FRQNT Doctoral Research Award, Physics Department Excellence Award and a Mitacs Globalink Graduate Fellowship.  R.K. is supported by ERC synergy grant nr. 610058 (``BlackHoleCam'').  C.J.L. acknowledges support from NSF grant 1611606.  D.M. is a Banting Fellow.  K.A. acknowledges support from NSF grant AAG-1714897.  B.A. is supported by a Chalk-Rowles Fellowship.  A.M.A. acknowledges funding from an NWO Veni fellowship.
    S.B.S. acknowledges support from NSF grant AAG-1714897.  F.K. thanks the Swedish Research Council. U.-L.P. receives support from the Ontario Research Fund Research Excellence Program (ORF-RE), NSERC, Simons Foundation, Thoth Technology Inc., and Alexander von Humboldt Foundation.  Z.Pl. is supported by a Schulich Graduate Fellowship.  P.S. is a Dunlap Fellow and an NSERC Postdoctoral Fellow.  The Dunlap Institute is funded through an endowment established by the David Dunlap family and the University of Toronto.  K.M.S. is supported by an NSERC Discovery Grant, an Ontario Early Researcher Award, and a CIFAR fellowship.  Part of this research was carried out at the Jet Propulsion Laboratory, California Institute of Technology, under a contract with the National Aeronautics and Space Administration. The NANOGrav project receives support from National Science Foundation (NSF) Physics Frontiers Center award number 1430284.  FRB research at UBC is supported by an NSERC Discovery Grant and by the Canadian Institute for Advanced Research.  The CHIME/FRB baseband system is funded in part by a Canada Foundation for Innovation John R. Evans Leaders Fund award to I.H.S. 
    The National Radio Astronomy Observatory (NRAO) is a facility of the National Science Foundation operated under cooperative agreement by Associated Universities, Inc.
    The Astronomical Image Processing System (AIPS) is a software package produced and maintained by NRAO.  The Common Astronomy Software Applications (CASA) package is software produced and maintained by NRAO.
    This research made use of APLpy, an open-source plotting package for Python hosted at \url{http://aplpy.github.com}, Astropy, a community-developed core Python package for Astronomy\cite{astropy2013}, and Matplotlib\cite{hunter2007}.

    \item[Author Contributions] B.M. is Principal Investigator of the EVN observing programme and led the radio imaging analysis of those data.  K.N. discovered the radio bursts and led the time-domain analysis.  J.W.T.H. guided the time-domain analysis and made major contributions to the writing of the manuscript.  S.P.T. and C.B. led the optical imaging and spectroscopic analyses.  Z.P. performed an independent analysis of the EVN data, and in previous years has pioneered the development of the EVN's capabilities for fast transient research.  A.K. created the software used to correlate the EVN baseband data.  M.B. analysed the chance coincidence probability.  R.K. arranged the Effelsberg PSRIX observations.  C.J.L. analysed the VLA imaging data.  D.M. determined the CHIME/FRB baseband position.  V.M.K. played a significant coordination role that enabled these results.  All other co-authors contributed to the CHIME/FRB discovery of the source or the interpretation of the analysis results and the final version of the manuscript.
    
    \item[Competing Interests] The authors declare that they have no competing financial interests.
    
    \item[Correspondence] Correspondence and requests for materials should be addressed to J.W.T.H.~(email: J.W.T.Hessels@uva.nl).
\end{addendum}

\clearpage
\newpage

\section*{Extended Data}

\setcounter{figure}{0}
\setcounter{table}{0}

\captionsetup[table]{name={\bf Extended Data Table}}

\begin{table}
 \caption{\label{tab:detection_properties}{\bf\frb burst properties as detected in both the Effelsberg PSRIX data and the Effelsberg auto-correlation data}. S/N and Width are determined using {\tt single\_pulse\_search.py}, and are the exact values returned by this program for the best-fit boxcar function. The DM quoted for the PSRIX data is the DM corresponding to the optimal S/N detection. The DM for the Effelsberg auto-correlation data is 348.76 pc cm$^{-3}$, which is the S/N optimising DM found for the brightest burst, B4, using the {\tt PSRCHIVE} tool {\tt pdmp}.\newline
 $^{\rm a}$ Time of arrival of the centre of the burst envelope at the Solar System Barycentre after correcting to infinite frequency (i.e. after removing the time delay from dispersion) using a DM of 348.76 pc cm$^{-3}$.}
\centering
\noindent
\resizebox{\textwidth}{!}{
\includegraphics{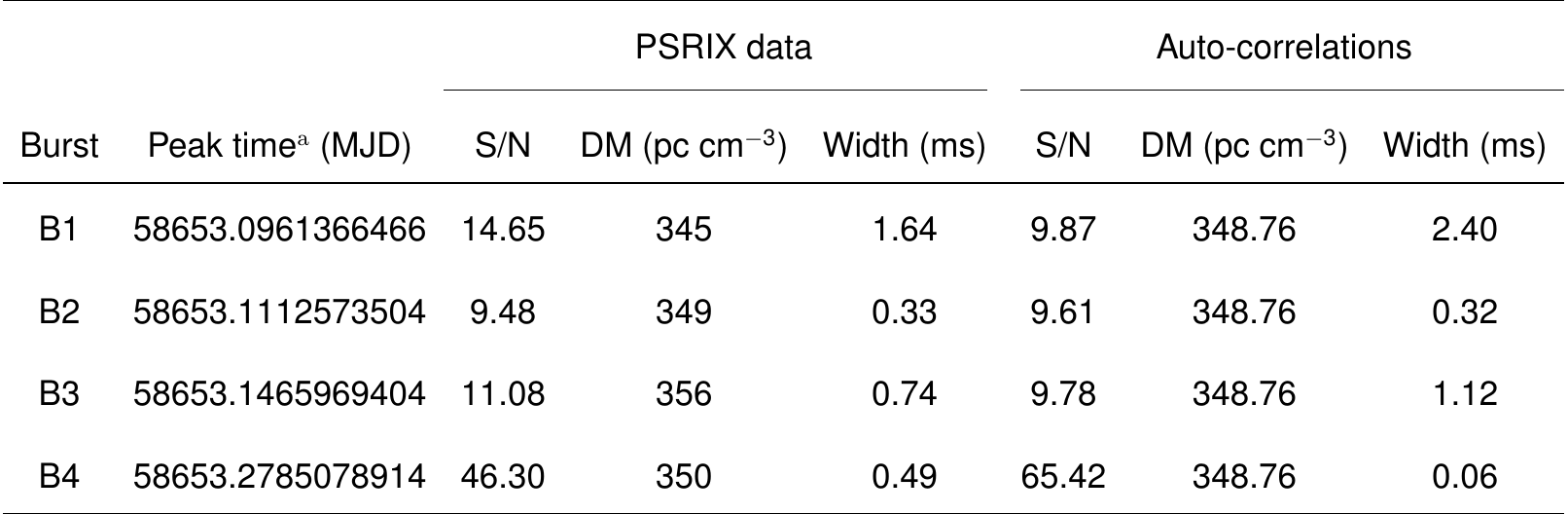}
}
\end{table}
\clearpage

\begin{figureED}
   \centerline{\resizebox{\textwidth}{!}{\includegraphics{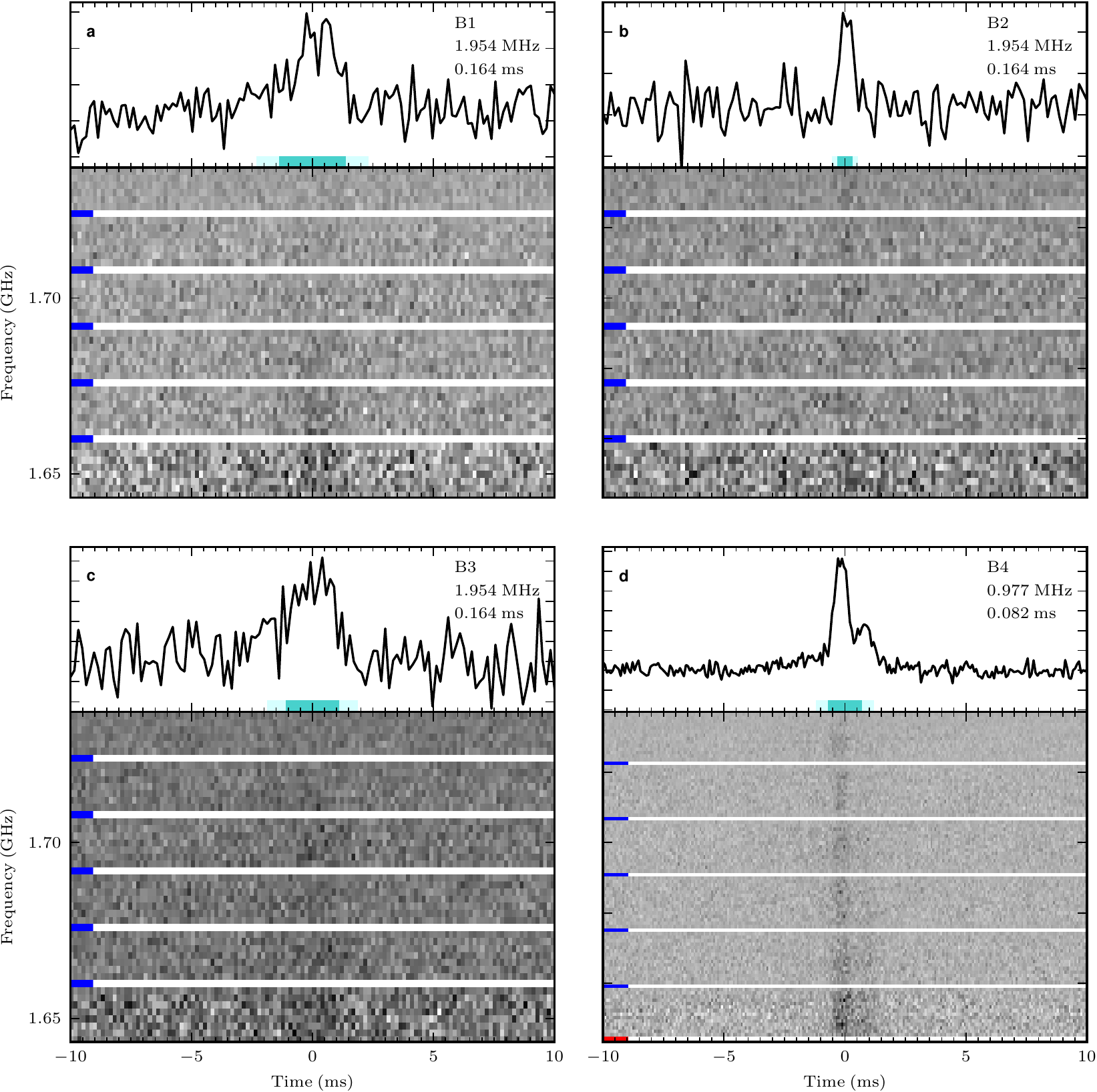}}}
    \caption{{\label{fig:efpsr_profile}
    {\bf Burst detections in Effelsberg PSRIX data.} 
    Band-averaged profiles and dynamic spectra of the four bursts, as detected in the PSRIX data ({\bf a}, {\bf b}, {\bf c} and {\bf d}). A 20-ms time window is shown surrounding the burst centre.  Each burst was fitted with a Gaussian distribution to determine the FWHM duration, which is represented by the cyan bars. The lighter cyan encloses the 2-$\sigma$ region.  The solid white lines are frequency channels that have been removed from the data due to either RFI or subband edges, indicated by the red and blue markers, respectively. For visual clarity, bursts B1, B2 and B3 ({\bf a}, {\bf b} and {\bf c}) are downsampled in both time and frequency by a factor of two.} The RFI excision was done prior to downsampling. The time and frequency resolution used for plotting is shown in the top right of each panel.}
\end{figureED}
\clearpage

\begin{figureED}
   \centerline{\resizebox{150mm}{!}{\includegraphics{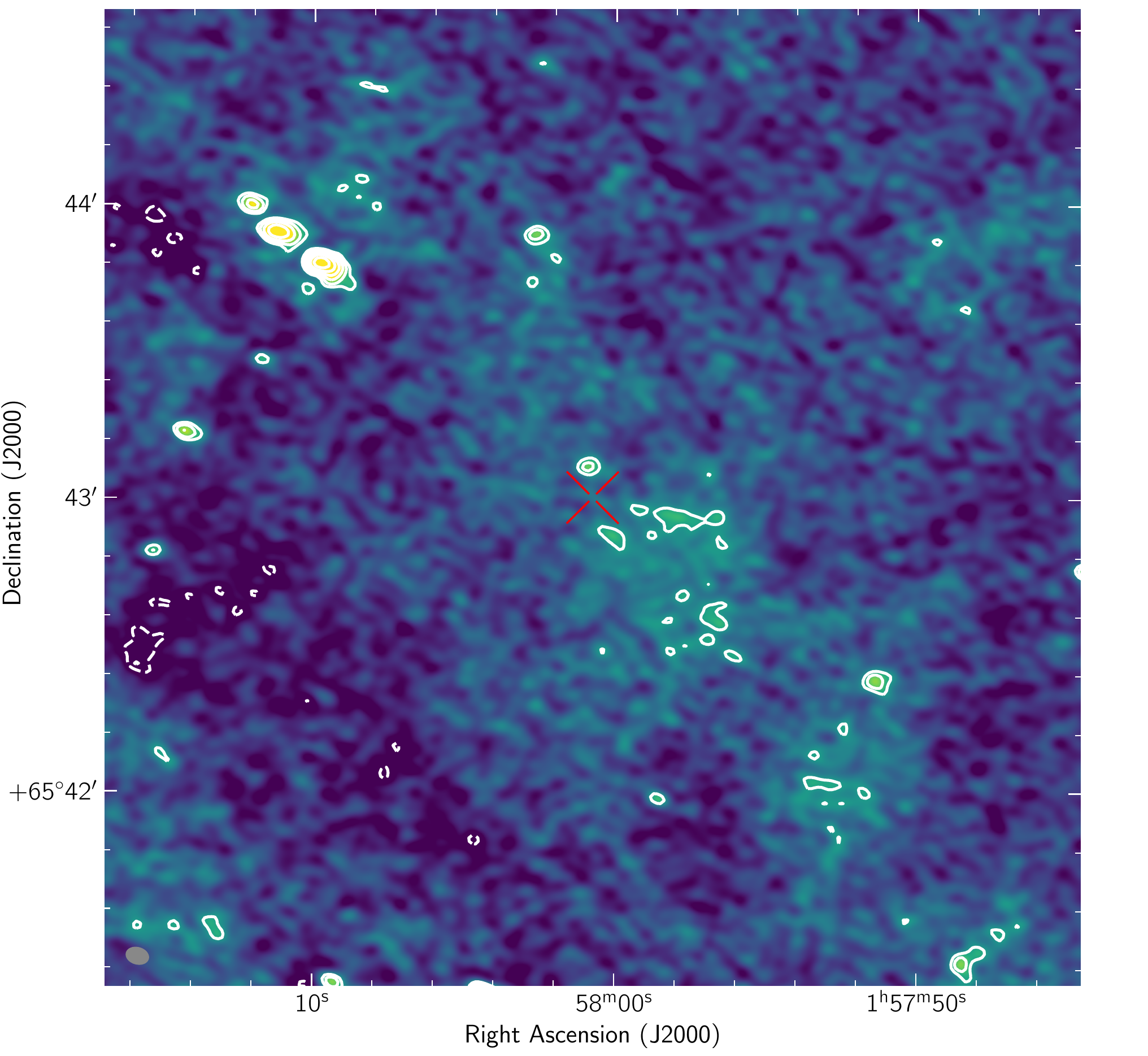}}}
    \caption{\label{fig:vla-map}
    {\bf VLA field image.}
    Field of the continuum radio emission around \frb as seen by the VLA at 1.6~GHz with a bandwidth of 0.6~GHz. The position of \frb is marked by the red cross at the centre of the image. Contours start at the 3-$\sigma$ r.m.s. noise level of 18~$\upmu$Jy~beam$^{-1}$ and increase by factors of $\sqrt{2}$. The synthesized beam is represented by the grey ellipse in the bottom-left corner. Note that a faint source is detected at around 6~arcsec north of \frb, but its separation is significant ($>$3-$\sigma$ confidence level) and we thus conclude that it is not associated with \frb.}
\end{figureED}
\clearpage

\begin{figureED}
    \centerline{\resizebox{150mm}{!}{\includegraphics{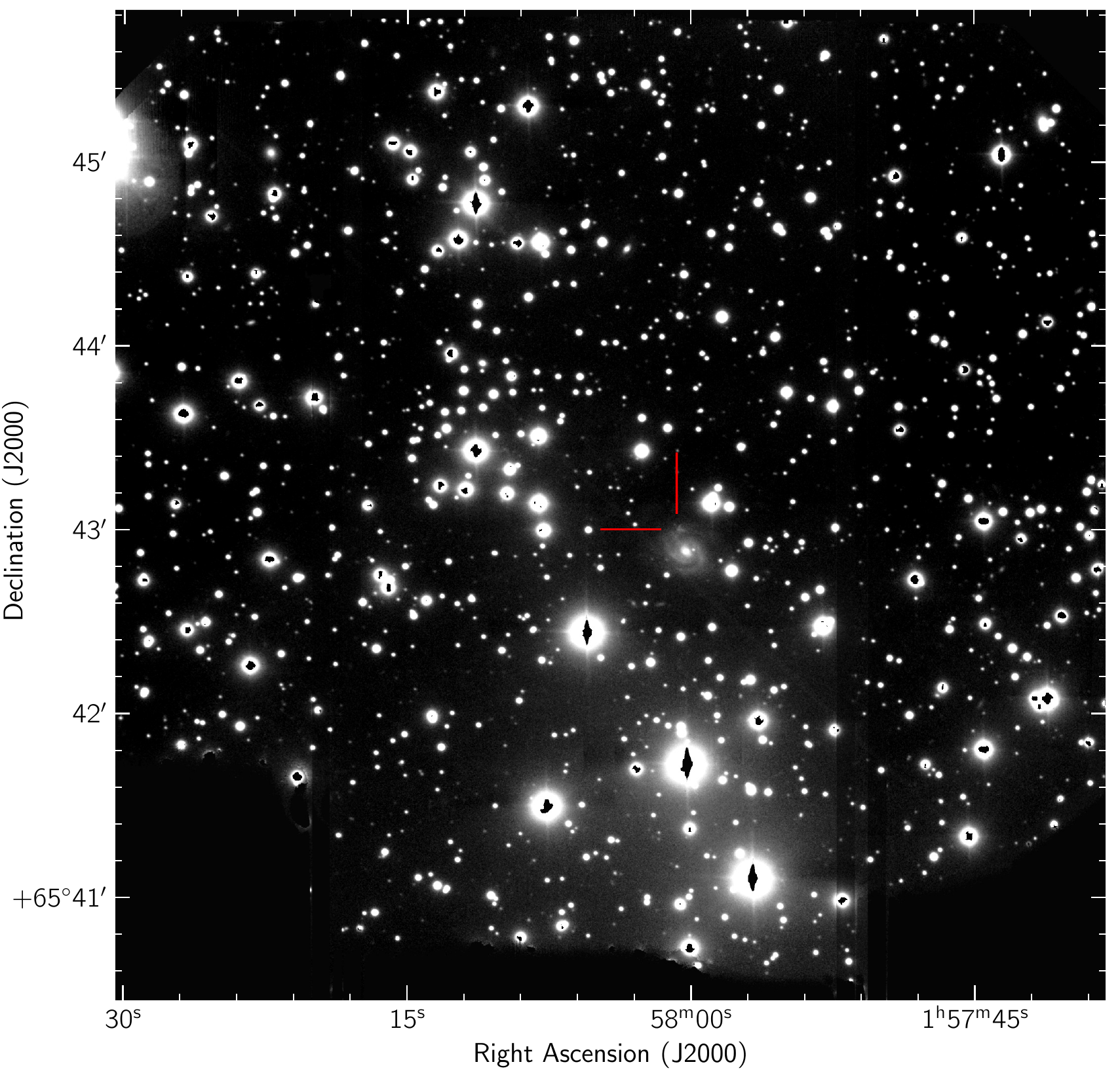}}}
    \caption{\label{fig:galaxy-wide-FoV}{\bf Full field of view of the Gemini $r^\prime$ filter}. The position of \frb is highlighted by the red cross. Note that the spiral galaxy associated with \frb is the only clearly visible galaxy in the field.}
\end{figureED}
\clearpage

\begin{figureED}
    \centerline{\resizebox{\textwidth}{!}{\includegraphics{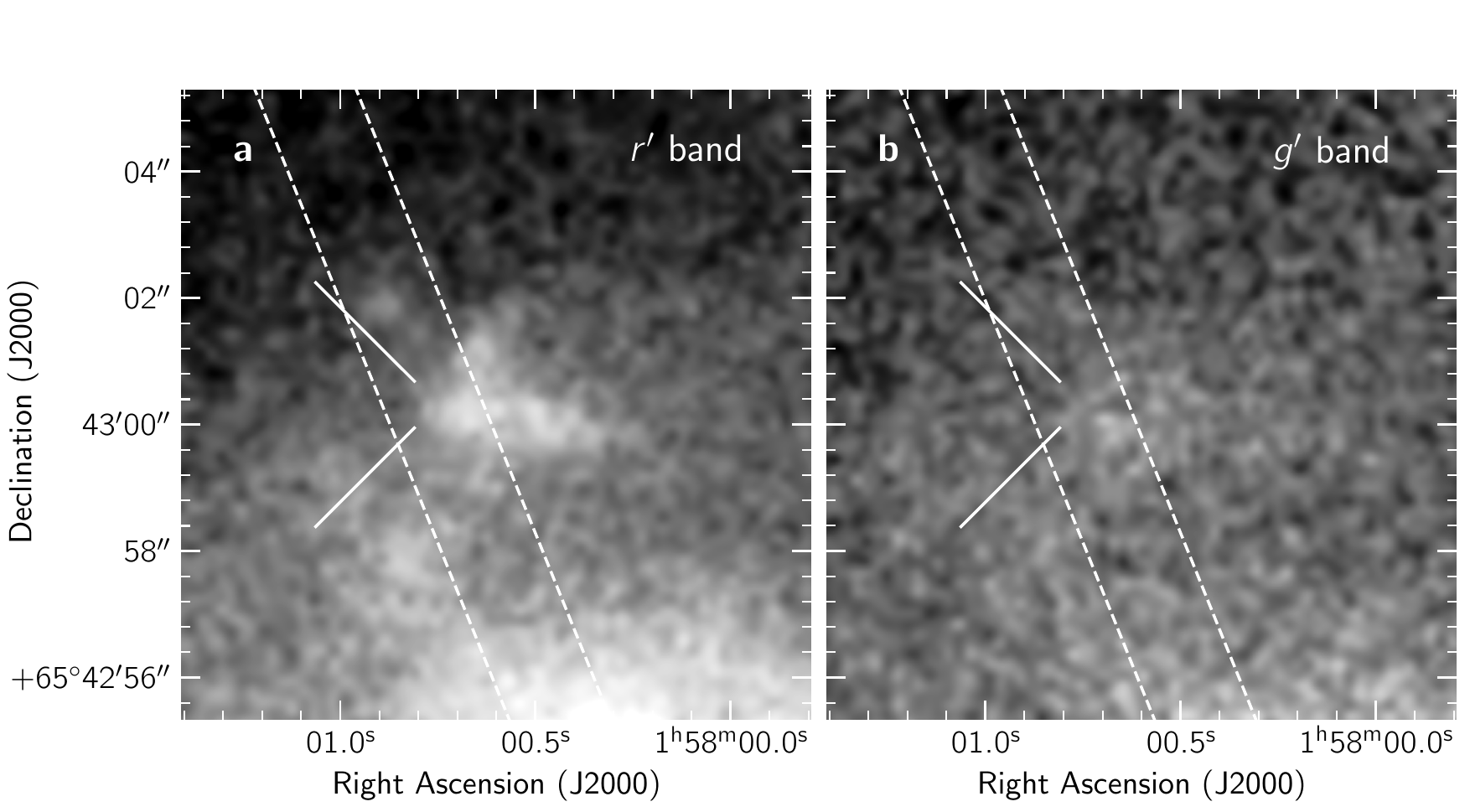}}}
    \caption{\label{fig:galaxy-zoom}{\bf Zoomed-in images at the position of \frb}.
    Gemini data at $r^{\prime}$ ({\bf a}) and $g^{\prime}$ bands ({\bf b}). The position of \frb is highlighted by the white cross. the uncertainty on its position is smaller than the resolution of these images. The dashed lines represent the orientation and placement of the 1.5~arcsec spectroscopic slit used to obtain the optical spectra. Note that the slit does not cover the full star-forming region but the region centred on \frb, and that the whole region is strongly affected by extinction ($E(g-r) = 1.73(9)$).}
\end{figureED}
\clearpage

\begin{figureED}
		\centerline{\resizebox{\textwidth}{!}{\includegraphics{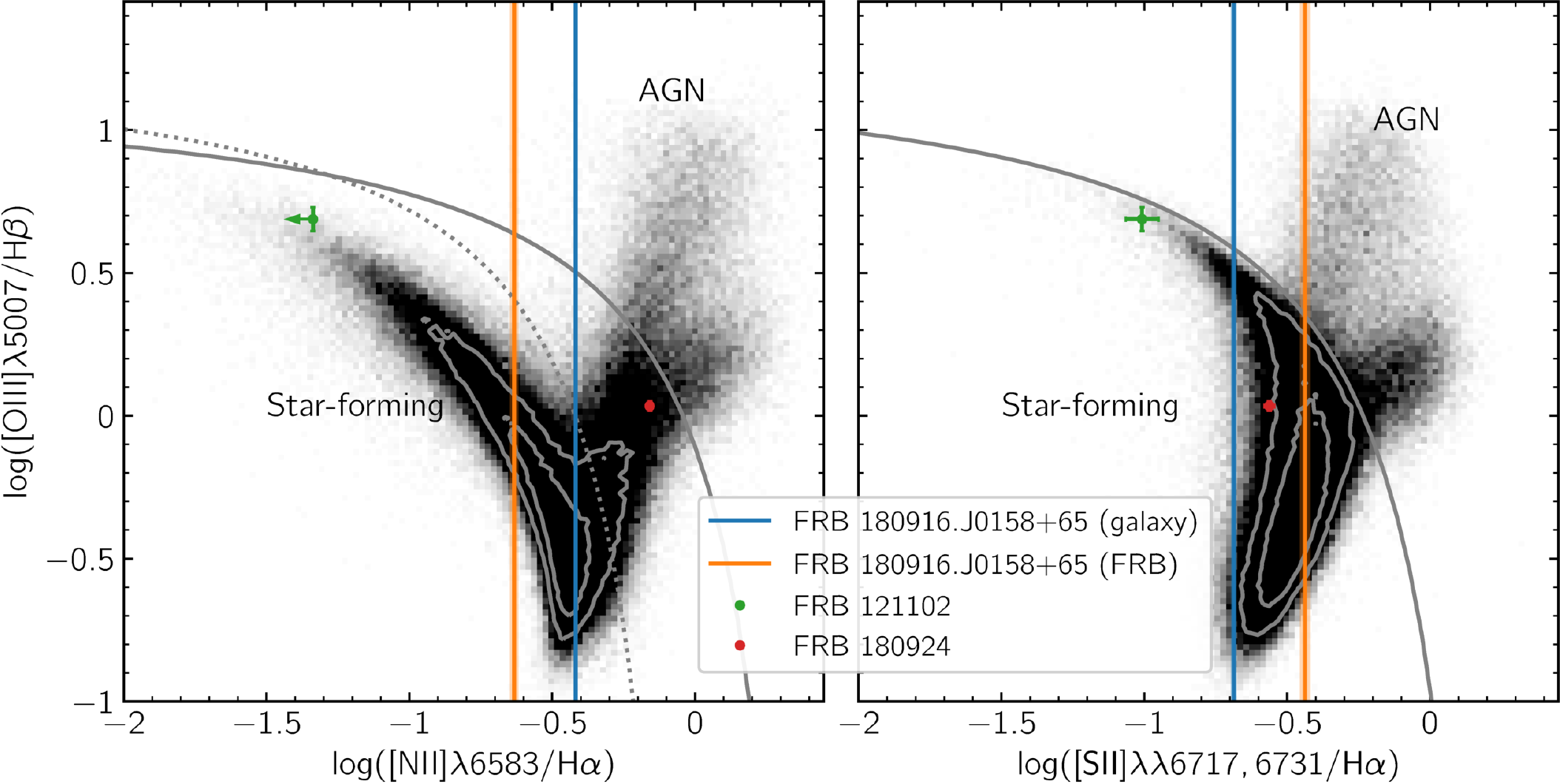}}}
		\caption{\label{fig:bpt} 
		{\bf Host galaxy source of ionisation.}
		Emission line flux ratios of {\bf a}: [\ion{N}{II}]/\Halpha and {\bf b}: [\ion{S}{II}]/\Halpha against [\ion{O}{III}]/H$\beta$. The grey-scale distribution represents the SDSS DR12 sample\cite{alam2015} of 240,000 galaxies that display significant emission lines ($>5\sigma$), where the solid and dotted grey lines denote the demarcations between star-forming and AGN-dominated galaxies\cite{kewley2001, kewley2002, kauffman2003}. The host galaxies of FRB~121102 and FRB~180924 are consistent with star-forming and AGN-dominated galaxies, respectively\cite{tendulkar2017,bannister2019}. 
		Though the Gemini-North spectrum of \frb does not cover the [\ion{O}{III}] and H$\beta$ lines, its [\ion{N}{II}]/\Halpha and [\ion{S}{II}]/\Halpha line ratios are broadly consistent with a star-formation dominated galaxy (represented by the vertical lines and the 1-$\sigma$ region as line width).}
\end{figureED}
\clearpage

\begin{figureED}
   \centerline{\resizebox{150mm}{!}{\includegraphics{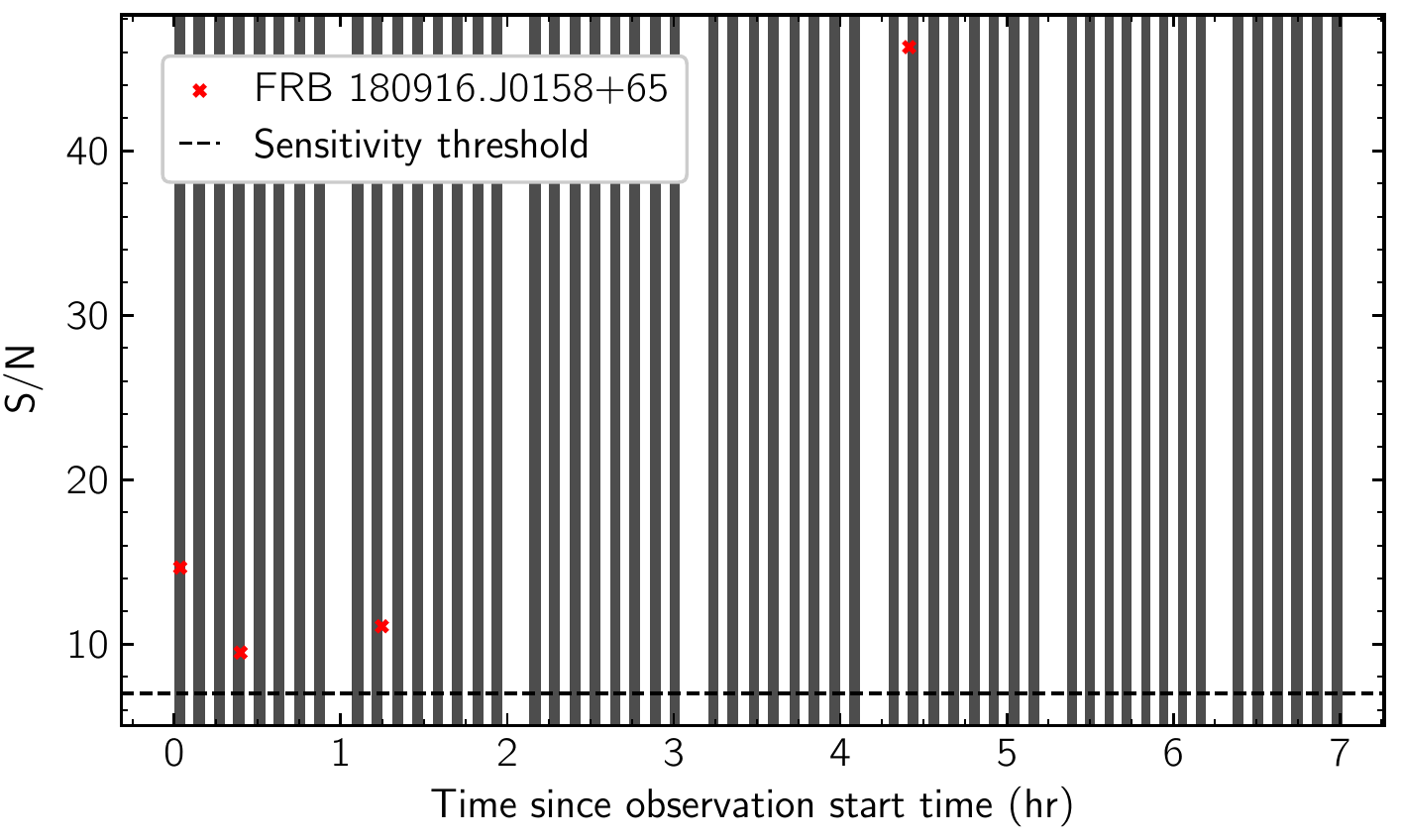}}}
    \caption{\label{fig:scans}
    {\bf Burst brightness and arrival times.}
    Burst S/N as a function of time during our 2019 June 19 observation of \frb. The grey bars represent scans of the \frb field. The red stars represent the four bursts (from left to right: B1, B2, B3, B4). The black dashed line indicates the detection threshold of our search in the pulsar-backend data (S/N = 7).}
\end{figureED}
\clearpage

\begin{figureED}
   \centerline{\resizebox{\textwidth}{!}{\includegraphics{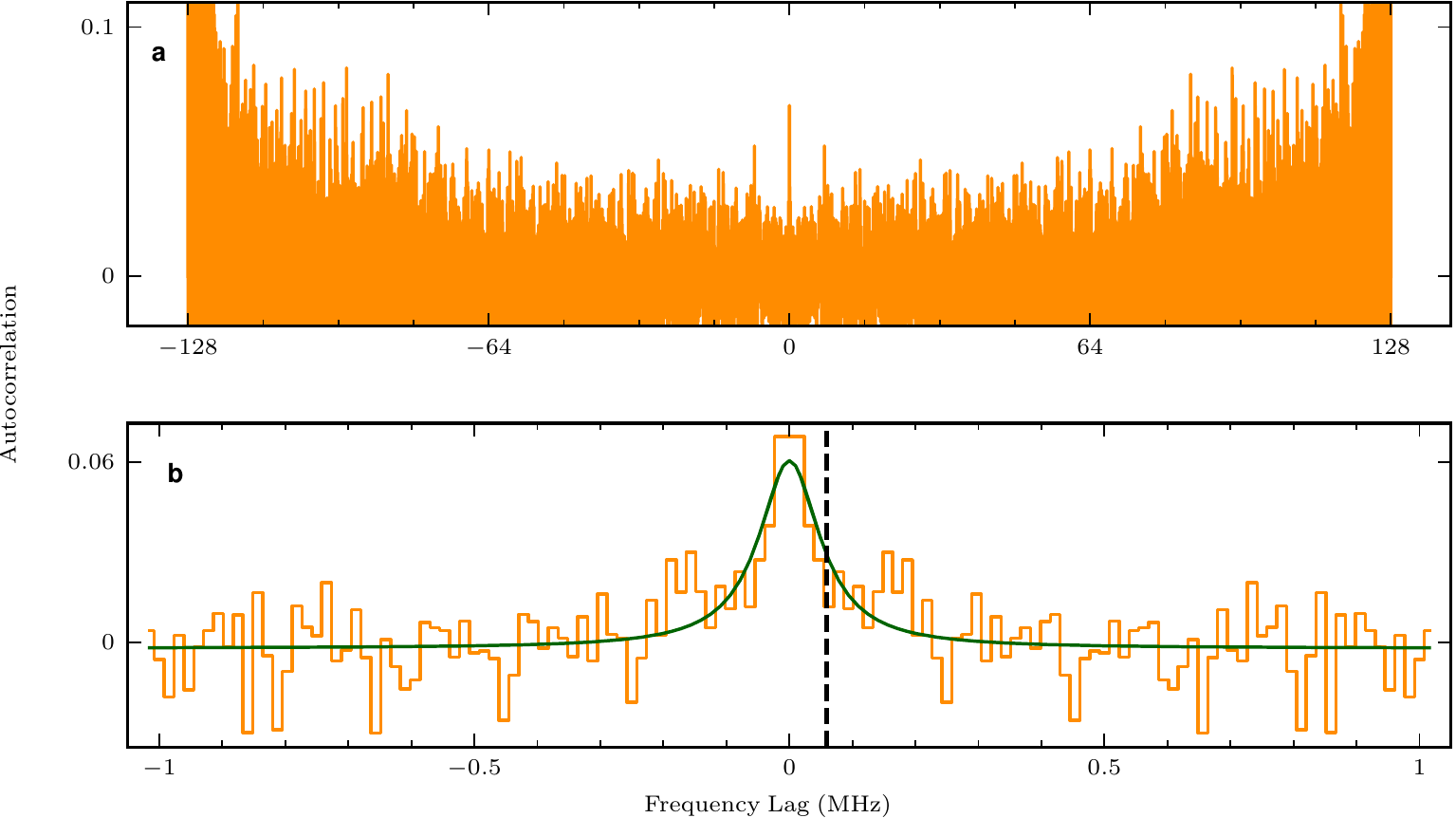}}}
    \caption{\label{fig:ACF}
    {\bf Auto-correlation function and scintillation bandwidth of brightest burst, B4.}
    {\bf a}: The auto-correlation function (ACF) of the spectrum of the bright, narrow burst component of burst B4. {\bf b}: The ACF for lags between $-$1.016 and 1.016 MHz. The zero-lag noise spike has been removed from the ACF. A Lorentzian fit is shown in green in panel ({\bf b}). The black vertical dashed line represents the scintillation bandwidth, defined as the half-width at half-maximum of the Lorentzian fit.}
\end{figureED}
\clearpage

\begin{figureED}
   \centerline{\resizebox{150mm}{!}{\includegraphics{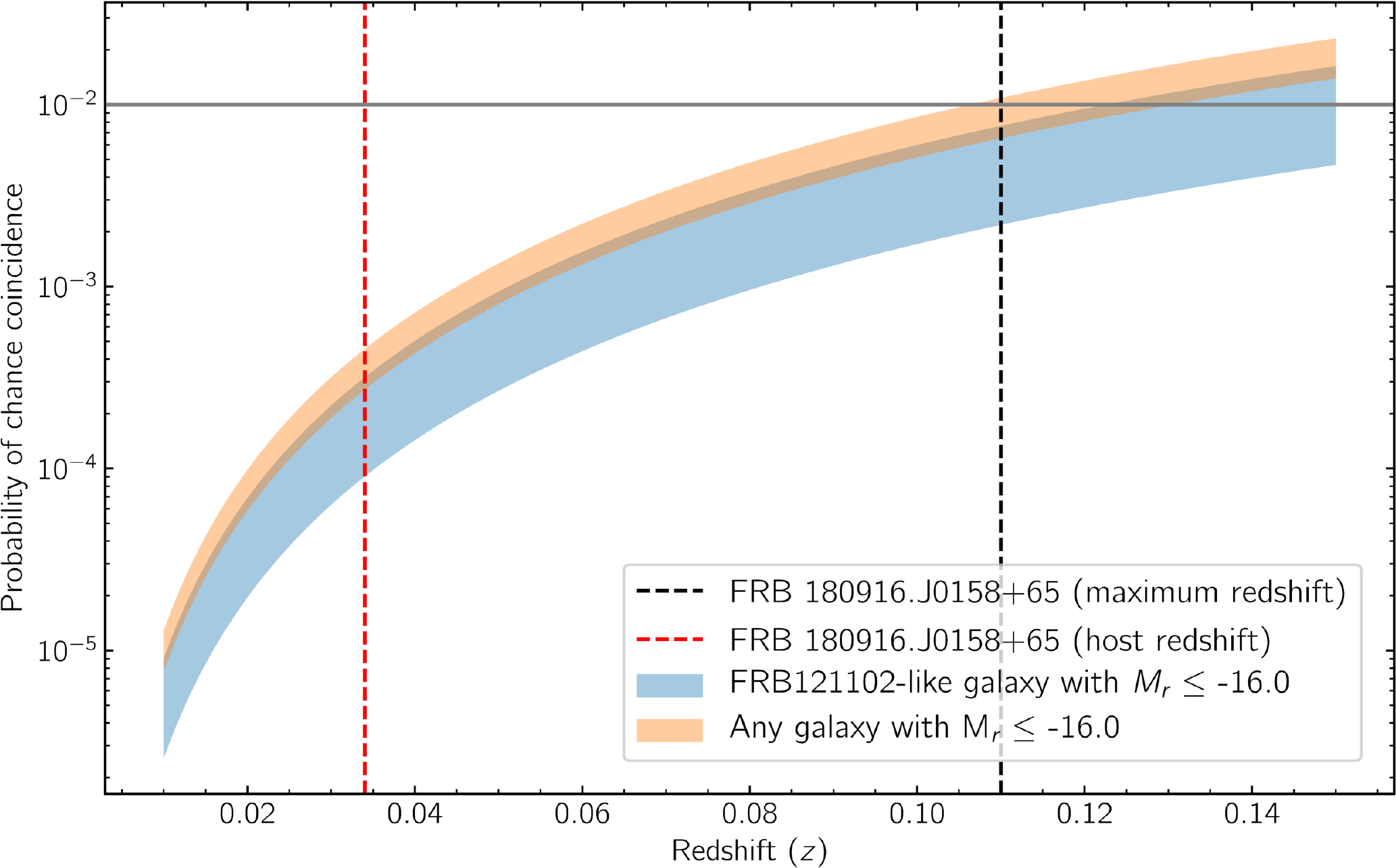}}}
    \caption{\label{fig:probability}
    {\bf Redshift-cumulated probability of chance alignment coincidence.}
    Probability of a chance alignment between \frb and twice the median half-light radius of any galaxy with magnitude $M_B \leq -16$ (orange region) or with a dwarf galaxy like the host of FRB~121102 (blue region) as a function of redshift. The horizontal grey line represents the 1\% probability threshold. At the redshift of the host galaxy, $z = 0.0337$ (vertical dashed red line), the chance coincidence probability is $P \ll 0.1\%$, and at the maximum possible redshift of $\sim 0.11$ derived from the observed DM the probability is $\lesssim 1\%$ (vertical dashed black line).}
\end{figureED}
\clearpage


\begin{table}
    \caption{\label{tab:spectral_lines}{\bf Properties of the spectral emission lines.} Properties of the most relevant spectral lines observed at the location of the core of the galaxy and at the position of \frb. Note that the fluxes are not corrected for extinction.  Numbers in parentheses indicate 1-$\sigma$ uncertainties in the least significant digits.}
    \medskip
    \resizebox{\textwidth}{!}{
    \includegraphics{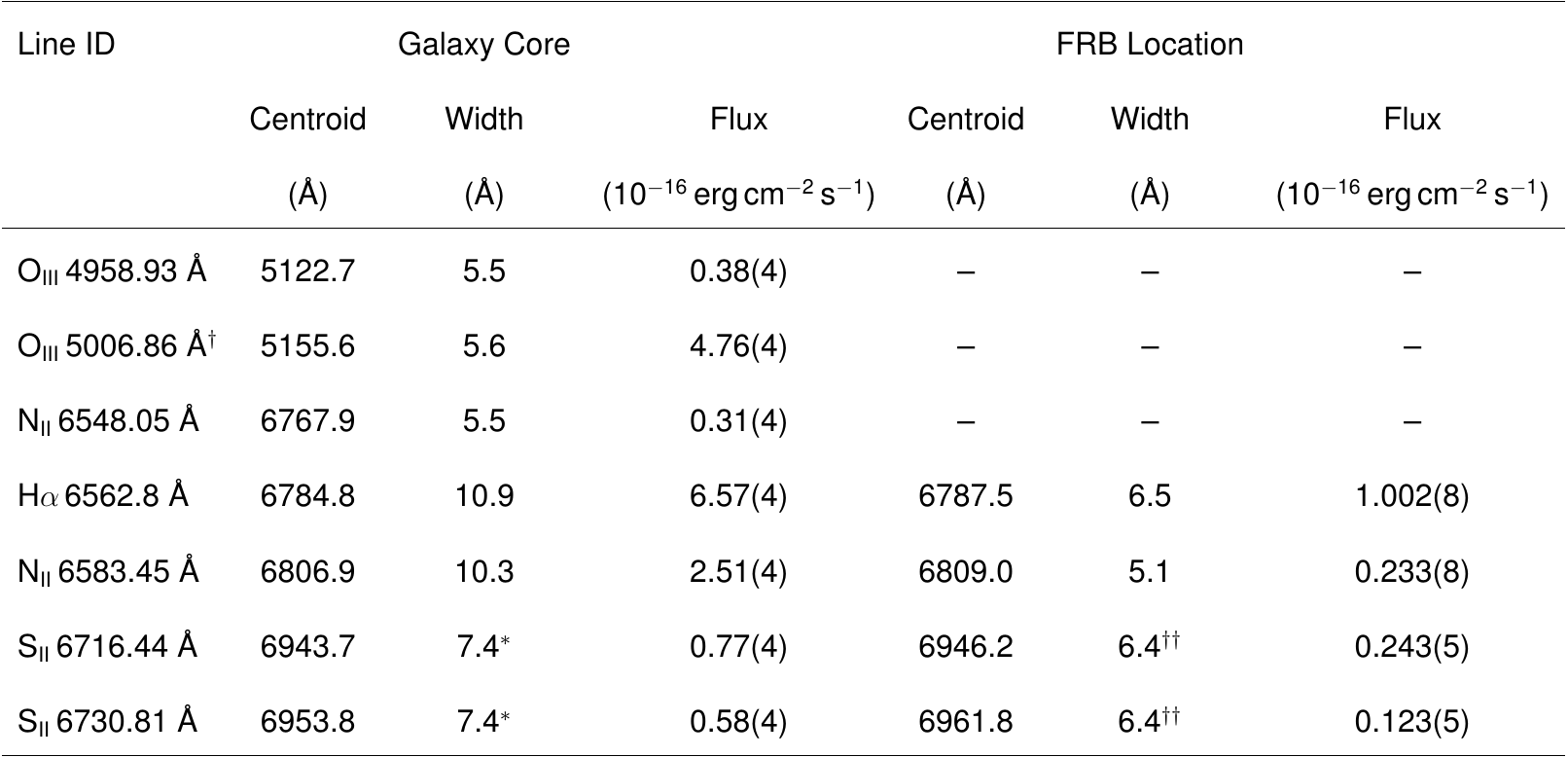}\\
    }
     Notes: $^{\dagger}$: The O$_\text{III}$\,5006.86~\AA\ line is detected at the FRB location but is corrupted by a cosmic-ray hit. $^{*}$,$^{\dagger\dagger}$: A single value of line width was fit for both lines. 
\end{table}

\end{document}